\documentclass[11pt,a4paper]{article}

\usepackage[T1]{fontenc}
\usepackage[utf8]{inputenc}

\usepackage{amsmath}
\usepackage{amsfonts}
\usepackage{amssymb}

\usepackage{fullpage}
\usepackage{enumerate}
\usepackage{setspace}
\numberwithin{equation}{section}
\usepackage[labelfont=bf]{caption}

\usepackage{color}
\definecolor{dark-gray}{gray}{0.20}
\definecolor{gray}{gray}{0.30}
\definecolor{light-gray}{gray}{0.80}
\definecolor{dark-red}{rgb}{0.7,0,0}
\definecolor{dark-green}{rgb}{0.1,0.4,0}
\definecolor{dark-blue}{rgb}{0.3,0.3,0.7}
\definecolor{light-blue}{rgb}{0.8,0.8,1}

\usepackage{cite}
\usepackage{hyperref}
\hypersetup{
	colorlinks=true,
	linkcolor=dark-blue,
	citecolor=dark-red,
	urlcolor=dark-green,
	linktoc=page
}

\usepackage{graphicx}
\usepackage[all,cmtip]{xy}

\renewcommand{\i}{\mathrm{i}}
\renewcommand{\d}{\mathrm{d}}

\newcommand{\dd}{\mathrm{d}}
\newcommand{\e}{\mathrm{e}}
\newcommand{\w}{\wedge}

\newcommand{\f}[2]{\frac{#1}{#2}}

\newcommand{\tr}{\text{tr}}

\renewcommand{\Re}{\text{Re}}
\renewcommand{\Im}{\text{Im}}
\newcommand{\Z}{{\mathbb Z}}
\newcommand{\C}{{\mathbb C}}
\newcommand{\CP}{{\C}P}
\newcommand{\RP}{{\R}P}

\newcommand{\R}{{\mathbb R}}
\newcommand{\T}{{\mathbb T}}

\renewcommand{\mod}{\,\, \text{mod} \,\,}
\newcommand{\beq}{\begin{equation}}
\newcommand{\eeq}{\end{equation}}
\newcommand{\bea}{\begin{eqnarray}}
\newcommand{\eea}{\end{eqnarray}}
\newcommand{\bbm}{\begin{bmatrix}}
\newcommand{\ebm}{\end{bmatrix}}
\newcommand{\nn}{\nonumber}
\newcommand{\fix}{\text{Fix}}
\newcommand{\sgn}{\text{sgn}}
\newcommand{\etal}{et al. }

\begin{document}

\renewcommand{\title}[1]{\vspace{5mm}\noindent{\LARGE\bfseries #1\par}\vspace{5mm}}
\newcommand{\authors}[1]{\noindent{\bfseries #1\par}\vspace{5mm}}
\newcommand{\address}[1]{\noindent{\itshape \small #1\par}\vspace{1mm}}
\newcommand{\email}[1]{\noindent\texttt{\scriptsize E-mail: #1}}

\begin{titlepage}
\begin{flushright}
\small ITP-UH-14/14\\
\normalsize
\end{flushright}
\vspace{2cm}

\newcommand{\HorRule}{\rule{\linewidth}{1pt}\\}
\begin{flushleft}\HorRule

\title{Wilson lines and Chern-Simons flux in explicit heterotic Calabi-Yau compactifications}
\authors{Fabio Apruzzi${}^{1,3}$, Fridrik Freyr Gautason${}^{1,2,4}$, Susha Parameswaran${}^{1,2}$ and\\Marco Zagermann${}^{1,2}$}
\address{${}^1$ Institut f\"ur Theoretische Physik \& ${}^2$ Center for Quantum Engineering and Spacetime Research, \\
$\,\,\,$ Leibniz Universit{\"a}t Hannover,
 Appelstra{\ss}e 2, 30167 Hannover, Germany \\
${}^{3}$ Department of Physics, Robeson Hall, Virginia Tech, Blacksburg, VA 24060, U.S.A. \\
${}^4$ Instituut voor Theoretische Fysica, K.U. Leuven, Celestijnenlaan 200D, B-3001 Leuven, Belgium}
\HorRule
\vspace{5mm}
\email{fabio.apruzzi@itp.uni-hannover.de, ffg@fys.kuleuven.be, susha.parameswaran@itp.uni-hannover.de and marco.zagermann@itp.uni-hannover.de }
\end{flushleft}
\vspace{2cm}

\begin{center}
\begin{minipage}{14cm}
We study to what extent Wilson lines in heterotic Calabi-Yau compactifications
lead to non-trivial $H$-flux via Chern-Simons terms. Wilson lines are basic
ingredients for Standard Model constructions but their induced $H$-flux may
affect the consistency of the leading order background geometry and of the
two-dimensional worldsheet theory. Moreover $H$-flux in heterotic
compactifications would play an important role for moduli stabilization and
could strongly constrain the supersymmetry breaking scale. We show how to
compute $H$-flux and the corresponding superpotential, given an explicit
complete intersection Calabi-Yau compactification and choice of Wilson lines.
We do so by identifying large classes of special Lagrangian submanifolds in the Calabi-Yau,
understanding how the Wilson lines project onto these submanifolds, and
computing their Chern-Simons invariants. We illustrate our procedure with
the quintic hypersurface as well as the split-bicubic, which can provide
a potentially realistic three generation model.
\end{minipage}
\end{center}
\end{titlepage}

\tableofcontents

\section{Introduction}
Heterotic string compactifications on Calabi-Yau (CY) manifolds with Wilson
lines have had considerable success in string model building
\cite{Braun:2005ux,Braun:2005nv, Bouchard:2005ag, Anderson:2009mh, Braun:2011ni, Anderson:2012yf,Braun:2013wr}, with abundant explicit examples
containing only a supersymmetric standard model, a hidden sector and a
few geometric and vector bundle moduli.  There are also several ideas on
how to address the moduli stabilization problem, although their realization
in explicit constructions has proven more challenging.  An important
observation is that the holomorphicity and
stability conditions on vector bundles could lift many of the flat directions
already at tree-level \cite{Witten:1985bz,Anderson:2010mh,Anderson:2011cza, Anderson:2011ty,Anderson:2013qca}.
Another mechanism proposed by \cite{Gukov:2003cy} is to stabilize moduli with
fractional $H$-flux sourced by Wilson lines in conjunction with gaugino
condensation. In ref. \cite{Cicoli:2013rwa} it was argued that 
this mechanism would generically lead to GUT scale supersymmetry
breaking.

Wilson lines were first introduced in order to break GUT gauge groups without breaking supersymmetry.  However, any concomitant
$H$-flux might also unintentionally affect the
self-consistency of the compactification background. Indeed, it is
well-known that the backreaction of $H$-flux deforms away from
supersymmetric Calabi-Yau compactifications of the leading order
10D heterotic supergravity theory, either by breaking supersymmetry or
by leading to non-K\"ahler internal spaces \cite{Strominger:1986uh}.
Moreover, it has also long been known that the Wilson lines' contribution
to $H$-flux may be associated with global worldsheet anomalies and could
thus be inconsistent as string backgrounds \cite{Witten:1985mj}.

Since for a given choice of Wilson lines and background manifold, the
fractional $H$-flux is completely determined and not a matter of choice, it
is important to develop techniques that allow one to compute it in concrete
examples to address the above issues. In this paper we focus on complete
intersection Calabi-Yau (CICY) manifolds (or rather quotients thereof by a
freely acting discrete symmetry group) as these provide a well understood class of 
potentially realistic
particle physics models \cite{Braun:2005ux,Braun:2005nv, Bouchard:2005ag,Anderson:2009mh, Braun:2011ni, Anderson:2012yf,Braun:2013wr}. In order to
compute the induced $H$-flux from given Wilson lines we use a class of special
Lagrangian submanifolds (sLags) as representatives of the three-cycles of
the CICYs. One reason for this is that these sLags are easily explicitly constructed
as fixed point loci of certain anti-holomorphic involutions that are
completely classified \cite{Partouche:2000uq}. Furthermore, the intersection
theory of sLags is particularly simple. We then show that the
projection of the Wilson line and its induced Chern-Simons term on these
sLags can be systematically determined. Hence, if the above
sLags span a basis for the third homology group (i.e. if the rank of their
intersection matrix matches the dimension of the third homology group),
the superpotential can be expressed as a linear combination of explicitly
computable Chern-Simons invariants on these sLags. Our procedure can then
be summarized as follows:
\begin{enumerate}
\item 
Identify 
sLags in the CICY under consideration, as fixed point sets of 
isometric anti-holomorphic involutions classified in
\cite{Partouche:2000uq}.  We do this in section \ref{involutions}.  Within this classification, we also show how the Wilson lines project onto the sLags in section \ref{S:WLonsLags}.
\item Calculate the intersection matrix of the sLags and compare its rank
with the dimension of the third homology group. We provide details and
further references on how this computation can be done systematically in
appendix \ref{S:appendix}.
\item 
Compute the Chern-Simons invariants
on the sLags. 
To this end we 
review some results from the
mathematics literature on Chern-Simons invariants on three-manifolds in
section \ref{CSonSeifert}. In order to apply these results one has to
determine the topology of the relevant sLags, and a central role will be played by Seifert fibered manifolds or compositions thereof.
\end{enumerate}

We begin the paper in section \ref{S:3formflux}, by recollecting some
well-known facts about $H$-flux in heterotic string compactifications. We
discuss the consistency of non-trivial $H$-flux, be it fundamental or induced
by Wilson lines, in supersymmetric CY compactifications, recalling subtleties
associated with the inclusion of gaugino condensation. On one hand, a
dimensional reduction of the 10D effective theory including non-trivial
$H$-flux and possibly fermionic bilinears does not allow for a supersymmetric
vacuum on CY internal spaces
\cite{Dine:1986vd, LopesCardoso:2003sp, Frey:2005zz}. On the other hand,
including non-perturbative effects together with threshold corrections
directly in the 4D effective theory, one can restore supersymmetry
\cite{Gukov:2003cy,Cicoli:2013rwa} in an anti-de Sitter vacuum. 
The 10D description of this 4D solution
is not yet understood \cite{LopesCardoso:2003sp, Frey:2005zz}. We
discuss the Chern-Simons contributions to $H$-flux from both non-standard
embeddings and Wilson lines. 
As we recall, Chern-Simons contributions from non-standard embeddings effectively correspond to higher
derivative corrections.  They preserve the leading order supersymmetric
CY compactification, and the would-be $\alpha'$-corrections to the 4D
superpotential vanish for the massless modes due to non-renormalization theorems
\cite{Witten:1985bz,Witten:1986kg, Green:1987mn}.
Wilson lines, in contrast, can contribute both to leading order $H$-flux and the
superpotential and are therefore potentially dangerous for the consistency
of the 10D solution. On a similar note, we also mention the relation between
$H$-flux due to Wilson lines and 2D global worldsheet anomalies
\cite{Witten:1985mj}.

In section \ref{S:CSflux} we give details on the procedure proposed above.
In section \ref{S:explicitmodels} and appendices \ref{S:appendix} and
\ref{A:CSSB}, we illustrate our method with two concrete models.
One of these is the special --- potentially realistic --- three generation
compactification on the quotient split-bicubic \cite{Braun:2005nv}.
We conclude the paper, in section \ref{S:conclusions}, with a summary and
discussion.

\section{The heterotic 3-form flux}
\label{S:3formflux}
In this section we will discuss two seemingly contradictory results that are important to bear in mind when considering $H$-flux in heterotic string compactifications.  Whether and how these results are concordant has not been worked out in detail.
\begin{itemize}
\item Compactifying leading order heterotic supergravity on CY 3-folds to a supersymmetric 4D maximally symmetric vacuum forces the 3-form flux $H$ to be zero. This is true even when vacuum expectation values of fermionic bilinears are taken into account in the 10D
action \cite{Dine:1986vd, LopesCardoso:2003sp, Frey:2005zz}.
\item By including the non-perturbative effects of fermionic condensates and threshold corrections 
directly in the effective
4D theory of a CY compactification, one can in principle turn on $H$-flux while simultaneously preserving
supersymmetry \cite{Gukov:2003cy}.
\end{itemize}
This section is therefore largely a review of the literature on various subtleties associated with $H$-flux and gaugino condensation on CY internal spaces. We will consider in particular the effects of non-trivial Chern-Simons terms in this context. We will also briefly discuss the 4D superpotential from Chern-Simons flux, 
considering the well known non-renormalization theorem. Finally, we will mention the relation between Chern-Simons flux and global anomalies in the associated 2D sigma model.

\subsection{Supersymmetry, $H$-flux and gaugino condensation}
\label{HgcCY}
The low energy effective action of the heterotic string written in the 10D
string frame takes the form \cite{Bergshoeff:1989de} (we use the conventions
of \cite{Green:1987mn})
\beq\label{action}
S = \int\e^{-2\phi}\star\ \left\{R + 4|\d\phi|^2 - \f{1}{2}|T|^2
-\f{\alpha'}{4}\tr(|F|^2 + 2\bar{\chi}D\chi)\right\}~,
\eeq
where $\phi$ is the dilaton, $R$ is the Ricci scalar, $F$ is the Yang-Mills
field strength, and $\chi$ is the gaugino.  Also, $T = H - \Sigma/2$, where
$H$ is the heterotic 3-form field strength, 
and the 3-form $\Sigma$ is the gaugino bilinear
\beq
\Sigma = \f{1}{24}\alpha' \tr(\bar{\chi}\Gamma_{MNR}\chi)\d x^{MNR},
\eeq
where $\Gamma_{MNR}$ is the antisymmetrization of three 10D $\Gamma$-matrices.
A supersymmetric solution of the action \eqref{action} requires the vanishing
of all supersymmetry variations, which for the dilatino $\lambda$, gaugino
$\chi$ and gravitino $\psi_M$, are \cite{Bergshoeff:1989de, Frey:2005zz}
\bea
\delta \lambda &=& -\f12 \partial_M\phi \Gamma^M\epsilon + \f{1}{24}\left(H_{MNR} + \f14\Sigma_{MNR}\right)\Gamma^{MNR}\epsilon,\label{dilatino}\\ 
\delta \chi &=& -\f14 F_{MN}\Gamma^{MN}\epsilon,\label{gaugino}\\
\delta\psi_M &=& \nabla_M\epsilon - \f18 H_{MNR}\Gamma^{NR}\epsilon + \f{1}{96}\Sigma_{NRS}\Gamma^{NRS}\Gamma_M \epsilon.\label{gravitino}
\eea
This system has been studied extensively in the literature (see e.g.
\cite{LopesCardoso:2003sp,Curio:2005ew,Manousselis:2005xa,Frey:2005zz,Lechtenfeld:2010dr,Chatzistavrakidis:2012qb}) for K\"ahler and non-K\"ahler internal spaces.  In
this paper, our focus will be on CY internal spaces, 
which is the most studied case.

$H$-flux in heterotic compactifications was discussed soon after the
foundational work on CY compactifications \cite{Candelas:1985en}.  The
seminal paper by Strominger \cite{Strominger:1986uh} showed that, for
supersymmetric Minkowski solutions, $H$-flux generates torsion and deforms
away from K{\"a}hlerity\footnote{CY compactifications with $H$-flux are, however, possible if we relax the condition of a maximally symmetric 4D external space and consider 4D domain wall solutions \cite{Klaput:2013nla}.}.  Indeed, the supersymmetry conditions imply
$H=*\d J$, so that the (3,0) and (0,3) contributions to $H$ must vanish, and the (1,2) and (2,1) contributions induce non-Kählerity.  
One question that has been considered is then what is the effect
of gaugino condensation on these statements, especially as the $H$-flux and
the fermion bilinear, $\Sigma$, corresponding to the 4D gaugino condensate,
appear in a related way in the 10D theory.

$H$-flux and gaugino condensation were first considered  in
\cite{Dine:1985rz,Derendinger:1985kk}.  For CY compactifications, the
vanishing of the gravitino variation together with the equations of motion
requires $\Sigma$ to vanish \cite{Dine:1985rz,Frey:2005zz}. The gaugino
condensate in 4D is expected to descend from a non-vanishing expectation value
of $\Sigma$. This would then imply that gaugino condensation is not compatible
with the supersymmetry conditions on CY internal spaces. 
However, $H$-flux and
gaugino condensation are compatible with a Minkowski $\times$ CY
compactification, if we allow supersymmetry to be broken spontaneously
\cite{Dine:1985rz}.  In detail, the condition for 4D Minkowski space fixes $T=0$,
which then leads to non-vanishing supersymmetry transformations for the
dilatino and part of the gravitino.   
Note that
satisfying the Minkowski condition $T=0$ requires balancing the
quantized $H$-flux against non-perturbative effects, which are exponentially
small at weak coupling \cite{Rohm:1985jv}. Dine \etal \cite{Dine:1985rz}
compared the scalar potential obtained from dimensional reduction with the
scalar potential obtained via a superpotential, $W\sim c + Ae^{-aS}$, directly
in 4D field theory.  The results matched up to power law corrections, which
had been neglected in the 10D analysis.

Gukov \etal in \cite{Gukov:2003cy} later argued from a 4D perspective that a
supersymmetric AdS solution is also possible with $H$-flux and gaugino
condensation, provided we include one-loop threshold corrections.  A
non-vanishing $H$-flux leads to the well known superpotential
\cite{Gukov:1999ya,Becker:2002jj}
\beq \label{supot}
W_{\text{flux}} = \int_{Y_3} H\w \Omega,
\eeq
where the internal space $Y_3$ is assumed to be a CY 3-fold with a holomorphic
3-form $\Omega$. When gaugino condensates are taken into account we also have
to include a corresponding term in the superpotential \cite{Shifman:1987ia}
\beq\label{gauginosupot}
W_\text{gaugino} \sim -\e^{-8\pi^2 f/C},
\eeq
where $f$ is the holomorphic gauge kinetic function of the gauge group from
which the gauginos condense and $C$ is the dual Coxeter number of the gauge
group. Gukov \etal \cite{Gukov:2003cy} showed that an AdS supersymmetric
solution is possible in the resulting 4D effective field theory provided that
threshold corrections are taken into account so that the gauge coupling
function takes the form
\beq
f= S+\beta T,
\eeq
where $S$ and $T$ are the dilaton and volume moduli, and $\beta T$ is the
one-loop correction term.  From this point of view, however, it is not
completely clear if the internal space can remain a CY 3-fold, as we lack
a 10D description of the 4D threshold corrections.  Also, as was alluded 
to earlier, the contribution to $H$-flux  from the Kalb-Ramond 2-form 
$\dd B$ is quantized to integers \cite{Rohm:1985jv}, which would imply that 
the dilaton is stabilized at strong coupling. However this problem is 
ameliorated by using the Chern-Simons contribution to $H$ from Wilson lines, which is only 
fractionally quantized \cite{Rohm:1985jv,Gukov:2003cy} and will be discussed in more detail in the following subsections.

An attempt to capture the 4D physics described above within the 10D theory was
made by Frey and Lippert in \cite{Frey:2005zz}, by solving the 10D
supersymmetry conditions.  However, as we have already seen, it is clear from
the leading order 10D equations that the internal manifold cannot be CY,
rather, the solutions they found were a product of 4D AdS spacetime and
non-complex internal spaces.   The treatment of higher order corrections in
the 10D theory which correspond to the gaugino condensates with threshold
effects in the 4D theory is still missing. In fact it is unclear how to
derive the full 4D superpotential from 10D in the presence of $H$-flux and
--- in particular --- gaugino condensation.  Usually, the 4D superpotential
can be derived from the gravitino supersymmetry variation.  But Frey and
Lippert \cite{Frey:2005zz} showed that the contributions from the fermion
bilinears $\Sigma$ (the gaugino condensate in 4D) cancel here, so that the
10D theory does not seem to catch the 4D non-perturbative effect (see also
\cite{LopesCardoso:2003af, LopesCardoso:2003sp}).

To summarize, if we have non-trivial $H$-flux together with a 10D fermion
bilinear, both non-supersymmetric Minkowski $\times$ CY compactifications
\cite{Dine:1985rz} and supersymmetric AdS $\times$ non-CY compactifications
\cite{Frey:2005zz,LopesCardoso:2003af,LopesCardoso:2003sp,Lechtenfeld:2010dr,Chatzistavrakidis:2012qb} are possible.  Matching these solutions to a corresponding solution obtained
directly in 4D (with gaugino condensates) is non-trivial and not fully
understood.  As for supersymmetric CY compactifications with non-trivial
$H$-flux and gaugino condensation, a 4D construction that also relies on
threshold effects was given in \cite{Gukov:2003cy} (see also
\cite{Cicoli:2013rwa}). 
A 10D construction of these solutions has so far not
been obtained, as -- at leading order -- $H$-flux and fermion bilinears in the equations of motion are not compatible
with vanishing supersymmetry transformations.

\subsection{The Chern-Simons flux}
For the heterotic string, the 3-form $H$, i.e. the gauge invariant field
strength for the Kalb-Ramond 2-form $B$, is given not  simply by $\d B$, but
rather as:
\beq\label{Hexpression}
H = \d B - \f{\alpha'}{4}\left(\omega_{3\text{Y}} - \omega_{3\text{L}}\right),
\eeq
where the 3-form $\omega_{3\text{Y}}$  is the Chern-Simons form
\beq
\omega_{3\text{Y}} = \tr\left(A\w F - \f{1}{3} A\w A\w A\right)~,
\eeq
which locally satisfies $\d \omega_{3\text{Y}} = \tr F\w F$, and similar
expressions can be written down for the Lorentz Chern-Simons form
$\omega_{3\text{L}}$.  The Bianchi identity for $H$ therefore has a non-trivial
contribution on the right hand side:
\beq\label{bianchi}
\d H = \f{\alpha'}{4}\left(\tr R\w R - \tr F\w F\right)\,,
\eeq
which requires $P_1(V;\R) = P_1(T;\R)$, that is, the first Pontryagin classes
over real numbers for the tangent bundle and vector bundle should be equal.  
It is important to note that the Kalb-Ramond and Yang-Mills Chern-Simons contributions are leading order in the derivative 
 expansion,
whereas the Lorentz Chern-Simons term is higher order. 
Anomaly cancellation and the integrated Bianchi identity, however, force both Chern-Simons 
contributions in \eqref{bianchi} to be effectively of the same order, and we will see below that some Yang-Mills contributions to $H$ are therefore suppressed. This suppression justifies the common notation (\ref{Hexpression}), where the
Yang-Mills contribution to $H$ is assigned the subleading order $\alpha'$, but, as we will discuss below,
there can also be Yang-Mills contributions at leading order not affected by anomaly cancellation.

 As a result of the Chern-Simons contributions to $H$, we can have a non-zero
$H$-flux, even if we choose $\d B=0$ globally.  The full expression for the $H$-flux superpotential is:
\beq
W = \int \left[\d B - \f{\alpha'}{4}\omega_{3\text{Y}}\right]\w\Omega \,.
\eeq
Note that the Lorentz Chern-Simons term in $H$ does not contribute to $W$ because it necessarily appears at higher order in the $\alpha'$ expansion, 
whilst the superpotential does not receive any perturbative corrections beyond the leading order term \cite{Dine:1986vd,Green:1987mn}.  We now 
consider the Yang-Mills Chern-Simons contribution to $W$.  The Yang-Mills Chern-Simons term in $H$ can give rise to a background $H$-field via 
both the non-standard embedding and Wilson lines.  These, however, affect the background solution and $W$ in different ways.  

 A (non-)standard embedding solves the leading order supersymmetry conditions using a holomorphic connection on a holomorphic stable vector bundle.  However, imposing also the 
leading order Bianchi identity, $\d H = -\f{\alpha'}{4} \tr F\w F$, would imply $F=0$ and  vanishing background gauge field \cite{Green:1987mn}.  The non-trivial gauge field and any torsion due to $H$-flux is induced only when balancing with the higher derivative effects, from the Lorentz Chern-Simons contribution, in the integrated Bianchi identity.  That is, in both the standard and non-standard-embeddings of spin connection into gauge connection, anomaly cancellation enforces that the Yang-Mills and Lorentz Chern-Simons contributions are effectively of the same -- higher -- order in the $\alpha'$ expansion.  The non-renormalization theorem then implies that $H$-flux due to the non-standard embedding does not contribute to\footnote{Note that
the non-renormalization theorem only applies to the light modes in the low energy effective field theory. A non-standard choice of holomorphic stable vector bundle in general fixes some of the would-be CY-moduli by obstructing the corresponding geometric deformations. Formal inclusion of these massive fluctuations in the low energy theory then does lead to a non-trivial $W$ for those modes  and reproduces their expected stabilization
 from a 4D point of view \cite{Witten:1985bz,Anderson:2010mh,Anderson:2011cza, Anderson:2011ty}.} 
 $W$.  Moreover, the non-renormalization theorem can then be used to argue that the non-standard embedding is a consistent solution to all 
 finite orders in perturbation theory \cite{Green:1987mn}.  Indeed, as $W=\d W=0$ in the background at leading order, this must remain true to all 
 finite orders, and there exists a supersymmetric 4D Minkowski solution.  The internal geometry is Calabi-Yau at leading order, and receives 
 corrections at higher order.  In contrast to the non-standard embedding, we will see next that Wilson lines are a wholly leading order effect, indeed they do not contribute to the integrated Bianchi identity.  
 A non-trivial $H$-flux  induced by Wilson lines may thus contribute to the background $W$, and spoil the consistency of the leading order 
 supersymmetric Calabi-Yau compactification.  Whether or not consistency can be restored by higher loop effects is an open question.

\subsection{Wilson lines}
Wilson lines are flat vector bundle connections, that is, non-trivial gauge
configurations with ${F}=0$ but a global restriction to setting ${A}=0$
everywhere.  In
particular, when the fundamental group of the CY is non-trivial, we can define
the gauge invariant Wilson line operator, which is an embedding of $\pi_1(Y_3)$
into the gauge group $G$:
\beq
{\rm WL}_{\gamma} = {\rm P} \,\text{exp}\left( \i \int_{\gamma} A^a T_a \right)\,,
\eeq
where $\gamma$ is a non-trivial homotopy cycle on the CY space, and
$\rm{P}\,\text{exp}$ denotes the path ordered exponential. As $b_1(Y_3)=0$,
there are no Wilson line moduli or corresponding \emph{continuous} Wilson
lines in CY compactifications. Instead we can have at most
\emph{discrete} Wilson lines corresponding to a finite fundamental group
on a CY.

Discrete Wilson lines were introduced into CY compactifications as a way to
break the gauge symmetry without breaking supersymmetry
\cite{Witten:1985xc,Green:1987mn}.  Indeed, since $F=0$, they do not contribute to the Yang-Mills supersymmetry equations.  
However, they may still contribute non-trivially to the other supersymmetry conditions and equations of motion via the
Chern-Simons term in $H$, eq. \eqref{Hexpression}.  Moreover, any $H$-flux and torsion induced by Wilson lines is leading order, as 
$A$ is non-trivial although $F=0$ exactly and is vanishing in the Bianchi identity.  Therefore, Wilson lines can
contribute to the background superpotential.  Notice that only the (0,3) and harmonic
part of $\omega_{3\text{Y}}$ contributes \cite{Cicoli:2013rwa}.

\subsection{Chern-Simons invariants and global worldsheet anomalies}
\label{S:CSI&anomalies}
The Chern-Simons contribution to the superpotential can be expressed in terms
of a Chern-Simons invariant.  Indeed, we can write
\beq
W = -\frac{\alpha'}{4} \int_{Y_3} \omega_{3\text{Y}} \w \Omega = -\frac{\alpha'}{4} \int_{\Lambda} \omega_{3\text{Y}}~,
\eeq
where $\Lambda$ is the 3-cycle Poincar\'e dual to the holomorphic 3-form $\Omega$.
In general, the Chern-Simons invariant cannot be computed directly, as an
expression for the gauge field is not known.  Indeed, the gauge field $A$ is
neither uniquely nor globally defined.

Chern-Simons invariants for flat vector bundles have been well-studied in the
mathematics literature.  In particular the Chern-Simons invariant
\beq
CS(A,Q) = \int_{Q} \omega_{3\text{Y}}~,
\eeq
has been computed explicitly for several real 3-dimensional manifolds, denoted here by $Q$.  In
section \ref{CSonSeifert}, we summarize the known results on Chern-Simons
invariants for a large class of real 3-manifolds. Among the simplest examples
that give a non-trivial Chern-Simons invariant are the Lens spaces
$S^3/\Z_p$ for which one obtains
\cite{Witten:1985mj,KirkKlassen,Rozansky:1993zx,Nishi:1998}:
\beq
CS(A,S^3/\Z_p) = -\sum_i \f{k_i^2}{2p} \mod \Z \label{CSLens}
\eeq
for a gauge connection $A$ with the Wilson line fitting into $SU(N)$ as
specified by the integers $k_i$,
\beq
\label{WLonLensSpace}
U = \text{diag}(\e^{2\pi\i k_1/p},\dots,\e^{2\pi\i k_N/p})~.
\eeq
It is obvious from this example that the Chern-Simons invariant can take
fractional values; in fact it is only defined modulo integers, as large gauge
transformations shift $CS(A,Q)$ by integer values.  
This is
precisely the reason why \cite{Gukov:2003cy} suggested to use Chern-Simons
flux instead of the integer quantized $\dd B$-flux for moduli stabilization as
it facilitates the balance between flux and non-perturbative effects at
weak coupling. This proposal has recently been discussed in a wider context in
ref. \cite{Cicoli:2013rwa}, where it was found that even the fractional
Chern-Simons flux would generically lead to GUT scale supersymmetry breaking.
From a phenomenological point of view, it is thus very important to know
whether a non-trivial Chern-Simons invariant is induced by a given set of
Wilson lines. This is also true for ensuring the mathematical self-consistency
of such a scenario, as the mutual consistency of unbroken supersymmetry,
internal CY geometry, and non-trivial 3-form flux could so far not be
rigorously established from a purely 10D or even a worldsheet point of view.
Regarding the consistency of the 2D theory the situation may be even more
demanding due to worldsheet anomalies that cannot be cancelled with any known
methods\footnote{In fact, the relationship between Wilson lines and global worldsheet anomalies was used in \cite{Witten:1985mj} to indirectly compute the Chern-Simons invariant on the Lens space.}. 
More specifically this case occurs when $CS(A,Q)$ is fractional for
a 3-manifold $Q$ that corresponds to a torsion class of $H_3(Y_3,\Z)$
\cite{Witten:1985mj,Gukov:2003cy}. Motivated by all this, it is the purpose
of the present paper to explicitly compute Chern-Simons invariants induced by
Wilson lines on a class of phenomenologically
realistic CY spaces.

\section{Computing Chern-Simons flux in explicit models}
\label{S:CSflux}
We will now proceed to develop a strategy to compute the Chern-Simons flux and
its superpotential for Calabi-Yau compactifications with Wilson lines, and
apply this strategy to some explicit models with promising phenomenology.
More concretely our focus is on complete intersection Calabi-Yau (CICY)
3-folds, which are common setups for model building in
\cite{Braun:2005ux,Braun:2005nv,Bouchard:2005ag, Anderson:2009mh, Anderson:2012yf,Braun:2013wr}.

\subsection{Quick introduction to CICY}
Here we sketch the relevant information from the vast literature on CICY
manifolds. Much more detailed discussion can be found in the pioneering papers
\cite{Green:1986ck,Candelas:1987kf} and in the textbook \cite{Hubsch:1992nu}.
A CY manifold may be constructed as the set of homogeneous solutions to a set
of polynomials determined by the configuration matrix
\beq
\left[\begin{array}{c|cccc}
\CP^{n_1} & m_{11} & m_{12} & \cdots & m_{1l}\\
\CP^{n_2} & m_{21} & m_{22} & \cdots & m_{2l}\\
\vdots    & \vdots &        & \ddots & \\
\CP^{n_k} & m_{k1} & m_{k2} & \cdots & m_{kl}
\end{array}\right].
\eeq
This matrix specifies a class of $l$ polynomials in the ambient space
\beq
\CP^{n_1}\times \CP^{n_2} \times \cdots \times \CP^{n_k}.
\eeq
We call each polynomial $P_i$, where $i=1,\dots,l$ corresponds to the $i$th
column of the configuration matrix, and the entries in the matrix specify
that each term in the $i$th polynomial must contain $m_{ji}$ powers of the
coordinates from $\CP^{n_j}$. The set of simultaneous homogeneous solutions
to all the polynomials is a compact and smooth K\"ahler subspace of the
ambient space provided that the polynomials are transverse, that is
$\d P_1\wedge \dots \wedge \d P_l \neq 0$ at all points of intersection,
$P_i=0$. The subspace is furthermore Ricci flat and therefore CY if
the configuration matrix satisfies
\beq
\sum_i m_{ji} = n_j + 1, \qquad \forall j=1, \ldots, k.
\eeq
Of course for each configuration matrix there are many different choices of
polynomials, most of which correspond to smooth CY manifolds. All smooth
complete intersections corresponding to the same configuration matrix are
diffeomorphic and therefore topologically equivalent as real manifolds.

All CICYs are simply connected, whereas model building requires multiply
connected CYs in order to allow GUT symmetry breaking by Wilson lines.
Multiply connected CYs can be obtained by quotienting a CICY by some
freely-acting discrete symmetry group $\Gamma$.  The fundamental group of the
quotient CICY is then non-trivial, $\pi_1(Y_3/\Gamma) = \Gamma$.  When
quotienting a given CICY configuration by $\Gamma$, one must of course
consider only polynomials that respect this symmetry. This significantly
lowers the dimensionality of the moduli space of the CY.

\subsection{Special Lagrangian submanifolds}
In order to compute the Chern-Simons fluxes in CY compactifications, we will need to construct explicit 3-cycles, which the fluxes thread.  
We will therefore consider special Lagrangian submanifolds (sLags), which provide explicit representatives of 3-cycles in a CICY space and 
moreover have a particularly simple intersection theory.  Slags in a CY space are real 3D submanifolds defined by the conditions:
\beq
J\; \vline_{\,Q} = 0 \quad \text{and} \quad \Im(\e^{\i \frac{\theta}{2}} \Omega) \vline_Q = 0\,,\label{slag}
\eeq
with $J$ the K\"ahler 2-form, and $\theta$ is the so-called calibration angle
associated with the sLag (see \cite{Joyce, Hitchin} for some introductory
lectures on these geometries).  They are volume minimizing in their homology
class, with the volume form given by
\beq
\Re(\e^{\i \frac{\theta}{2}}\Omega) \vline_Q = \d\textrm{Vol}_Q \,.\label{calibrated}
\eeq
  Although general sLag submanifolds are difficult to construct explicitly,
there is one well-known method to obtain examples.  An isometric
anti-holomorphic involution\footnote{The isometricity property is
$\sigma(g)=g$, whereas anti-holomorphicity is $\sigma(I)=-I$ for $I$ the
complex structure.  Also $J=Ig$, and $g, J$ only define $\Omega$ up to a phase,
$J\wedge J\wedge J = \frac34 i \Omega \wedge \bar\Omega$.} $\sigma$ acts on the
CY manifold as
\beq
\sigma(J)=-J\, \quad \sigma(\Omega)=\overline{\e^{\i \theta}\Omega} \,.\label{antiholinv}
\eeq
Therefore, the fixed locus of $\sigma$ is a sLag submanifold; we will write
this as
\beq
Q_\sigma = \fix(\sigma) \,,
\eeq
where $Q_\sigma$ is the sLag and $\fix(\sigma)$ denotes the fixed point set of
the involution $\sigma$. Given a CICY with defining polynomials $P_i$, an
isometric anti-holomorphic  involution $\sigma$ on the ambient space descends
to the CICY if it satisfies
\beq\label{involutioncondition}
P_i\circ \sigma = \bar{P_i}.
\eeq

The sLag submanifolds in a CICY are therefore 3D submanifolds and
give rise to 3-cycles, which we can construct and analyze explicitly
using the defining polynomials.  As we will see in appendix \ref{S:appendix},
their intersection theory is also simple, so that it is straightforward
to check whether a given set of sLags generates the full third homology group
of the CICY. Furthermore, all the information required can be obtained by
going to a simple point in moduli space, that is, choosing a particularly
symmetric form of the defining polynomials, for which we can find many
homologically distinct sLags. Let $Q_\sigma$ be one such sLag.  As mentioned,
different polynomials corresponding to the same configuration matrix
determine manifolds that are diffeomorphic, so if $\tilde Y_3$ is another CICY
corresponding to the same configuration matrix as $Y_3$, then there
exists a diffeomorphism $f$ between $Y_3$ and $\tilde Y_3$. The restriction of
$f$ to $Q_\sigma$ defines a submanifold $f(Q_\sigma)$ in $\tilde Y_3$, which
may or may not be a sLag (in fact, sLags turn out to be surprisingly stable
under deformations of the CY structure \cite{Joyce}). As we are interested
in topological properties of the sLags as representatives of their homology
class, namely their Chern-Simons invariants, our final results will be
independent of these choices.

\subsection{A classification of sLags in CICYs}\label{involutions}
We will now provide a classification of the sLags in CICYs, which correspond
to the fixed point sets of isometric anti-holomorphic involutions. We will
start with relevant involutions on the ambient space; these will descend
to the CICY when the condition \eqref{involutioncondition} is satisfied.
Isometric anti-holomorphic involutions on $\CP^n$ can be classified into two
different types, $A$ and $B$ which act on the coordinates in the following
way \cite{Partouche:2000uq}
\bea
\sigma_A: (z_1,z_2,\dots,z_n,z_{n+1}) &\mapsto& (\bar{z}_1,\bar{z}_2,\dots,\bar{z}_n,\bar{z}_{n+1}), \label{sigmaA}\\
\sigma_B: (z_1,z_2,\dots,z_n,z_{n+1}) &\mapsto& (-\bar{z}_2,\bar{z}_1,\dots,-\bar{z}_{n+1},\bar{z}_n)\,.
\eea
Note that $\sigma_B$ applies only for projective spaces $\CP^n$ with $n$ odd.
All other involutions of $\CP^n$ can be constructed by a projective
$GL(n+1,\C)$ transformation acting on either $\sigma_A$ or $\sigma_B$
\cite{Partouche:2000uq},
\beq
\sigma_{A,B}^U = U^{-1}\circ \sigma_{A,B} \circ U~. \label{ABUDef}
\eeq
We will use the terminology $A$($B$)-type involution for an involution that
is constructed by the action of $GL(n+1,\C)$ on $\sigma_A$($\sigma_B$).
Note that $B$-type involutions act freely on $\CP^n$ and  therefore
$\fix(\sigma^U_B)$ is empty for all $GL(n+1,\C)$ transformations $U$. For
the $A$-type involutions, $\fix(\sigma_a^U)$ is non-empty and furthermore
\bea \label{eq:fixU}
\fix(\sigma_A^U) &=&\{z\in \CP^n \,|\, \sigma_A^U(z)=z\}  \nonumber \\
                 &=& \{z\in \CP^n \,|\, U^{-1}\overline{Uz}=z\}  \nonumber \\
                 &=& U^{-1}\{(z'=Uz)\in \CP^n \,|\, \overline{z'}=z'\} \nonumber\\
                 &=& U^{-1}\fix(\sigma_A) ~.
\eea
Applying this to a CY hypersurface in $\CP^n$, we see that if $\sigma_A$ is an
involution on the CY, then all matrices $U$ that are symmetries of the defining
polynomial will give involutions $\sigma_A^U$ on the CICY, and the corresponding
sLags are
\beq
Q_{\sigma_A^U} = U^{-1}(Q_{\sigma_A})~.
\eeq
This is an important
result that, in particular, shows that all $A$-type sLags are homeomorphic
$Q_{\sigma_A^U} \sim Q_{\sigma_A}$.

In the following, it will sometimes be useful to write the A-type involutions in terms of the matrices $M\equiv U^{-1}\overline{U}$, 
\begin{equation}
\sigma_{A}^{U}= M\circ \sigma_A .
\end{equation}
 
The $A$-type involutions on $\CP^n$ generalize to products of projective
spaces, for which the basic $A$-type involutions act individually on each
factor with complex conjugation
\beq
(\sigma_A,\sigma_A,\dots,\sigma_A):\CP^{n_1}\times \CP^{n_2}\times \cdots \times \CP^{n_k}\to\CP^{n_1}\times \CP^{n_2}\times \cdots \times \CP^{n_k}~.
\eeq
The fixed point set is given by
\beq
\fix(\sigma_A,\sigma_A,\dots,\sigma_A) = \RP^{n_1}\times \RP^{n_2}\times \cdots \times \RP^{n_k}~.
\eeq
A general $A$-type involution is now given by the map
\beq
(M_1 \circ \sigma_A,\dots,M_k \circ \sigma_A):\CP^{n_1}\times  \cdots \times \CP^{n_k}\to\CP^{n_1} \times \cdots \times \CP^{n_k}~.
\eeq
where the matrices $M_1,\dots,M_k$ are given in terms of $GL(n_i+1,\C)$
transformations $M_i = U_i^{-1}\overline{U}_i$. The fixed point set in this case
is given by
\beq
(U_1^{-1},U_2^{-1},\cdots,U_k^{-1})\fix(\sigma_A,\sigma_A,\dots,\sigma_A)~.
\eeq
In this paper we will only make use of diagonal matrices $U$ to generate sLags,
and the condition \eqref{involutioncondition} will then often force the diagonal
elements to be roots of unity.

When we have a product space of two identical projective spaces
$\CP^n\times \CP^n$ there is another type of involution, which we will call
$C$ \cite{Partouche:2000uq}:
\beq
\sigma_C:(z_i,w_i) \mapsto(\bar{w}_i,\bar{z}_i)~. \label{sigmaC}
\eeq
It is easy to see that the fixed point set of $\sigma_C$ is the diagonal in
$\CP^n\times \CP^n$,
\beq \label{fixsigmaC}
\fix(\sigma_C) = \{(z,\bar{z}) \in \CP^n\times\CP^n\}\,.
\eeq
All $C$-type involutions can be constructed by a pair of $GL(n+1,\C)$
transformations $U_1$ and $U_2$
\beq \label{UsigmaC}
\left(M,\;\overline{M}^{-1}\right) \circ \sigma_C:\CP^{n}\times \CP^{n}\to\CP^{n} \times \CP^{n}~,
\eeq
where $M = U_1^{-1}\overline{U}_2$ and the fixed point set is found to be
\beq
\left(U_1^{-1},\; U_2^{-1}\right)  \, \fix(\sigma_C)\,.
\eeq
Therefore, assuming that $(U_1,U_2)$  is a symmetry of the defining polynomials of the CICY, it gives rise to a sLag $Q_{\sigma_C^{(U_1,U_2)}}$, which is homeomorphic to the basic $C$-type sLag $Q_{\sigma_C}$.  Here, as for the
$A$-type sLags, we will restrict our attention to diagonal matrices $U_1$ and
$U_2$ which by \eqref{involutioncondition} usually forces the elements to be
roots of unity.

Having identified sLags via the isometric anti-holomorphic involutions of the
CICY, an important question will be how the quotient symmetry $\Gamma$, which
is freely acting on the CICY, acts on the sLags.  We will now turn to this and
related questions.

\subsection{Wilson lines on sLags}
\label{S:WLonsLags}
Our objective is to compute the contribution from discrete Wilson lines to the
Chern-Simons invariant on a given sLag. Consider a field $\phi$ on a quotient
CY, $Y_3/\Gamma$, transforming in some non-trivial representation of the GUT
gauge group. 
Each element, ${\mathsf g}$, of the fundamental group, $\Gamma$, of $Y_3/\Gamma$ defines
an action of $\Gamma$ on $\phi$ by parallel transport with respect to the gauge connection,
\beq
{\mathsf g}:\phi \mapsto {\rm WL}_{\mathsf g} \cdot \phi~,
\eeq
where
\beq
{\rm WL}_{\mathsf g} = \textrm{P}\,  \text{exp}\left( \i \int_{\gamma_{\mathsf g}} A^a T_a  \right)~
\eeq
is the Wilson line operator with a homotopy loop $\gamma_{\mathsf g}$ corresponding to
${\mathsf g}$, and the dot refers to the action on $\phi$ induced by its gauge group
representation. Without Wilson lines, this action is of course trivial.
As the fundamental group $\Gamma$ is discrete and the Wilson line operators
define a group homorphism, it is sufficient to specify the Wilson line
operators, WL$_{g}$, corresponding to the \emph{generators}, $g$, of $\Gamma$.

\begin{figure}[ht!]
\centering
\includegraphics[scale=0.5, angle=90, clip=true, trim=6cm 0cm 6cm 0cm]{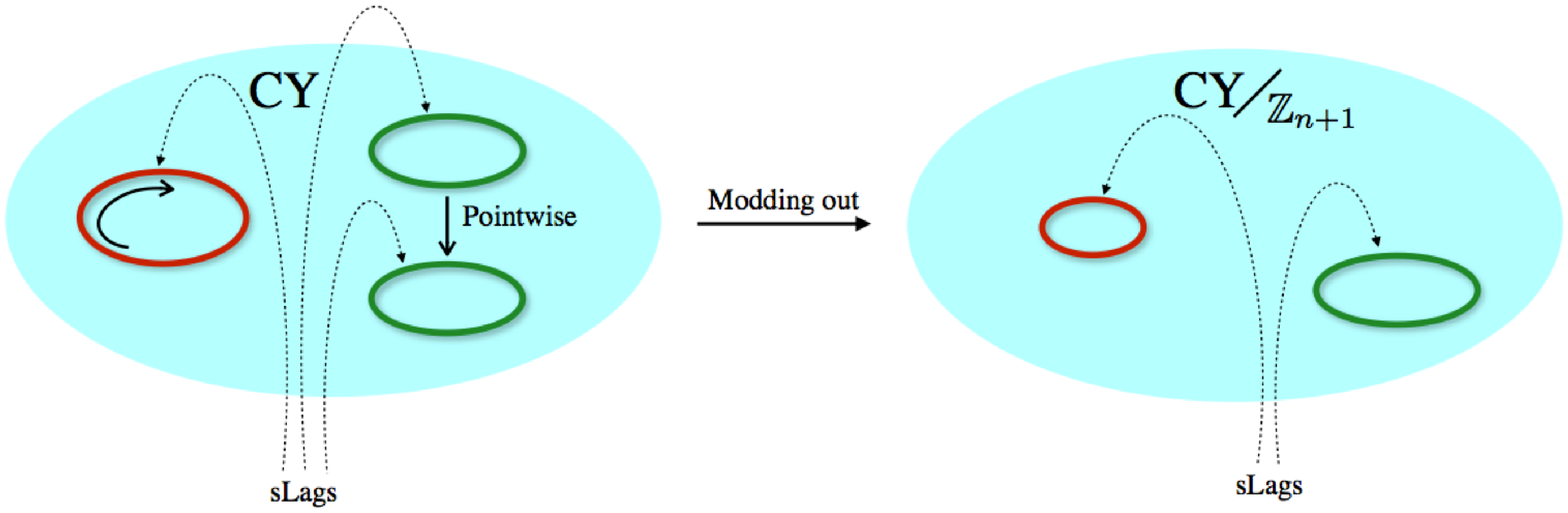}
\caption{A cartoon of the two possibilities for the free action of a
generator of $\Gamma$ on sLags. The black arrows depict
the action of the generator. On the left hand side we show the action on the covering CY.  It can act freely within the sLag (red), or
it can identify two (or more) distinct ones (green).
On the right hand side we can see what happens in the quotient CY. The red
sLag can have modified topology, because of the free action of the
generator. The green sLags are simply identified, and the resulting sLags
have the same topology as before. Only the red sLags can possibly inherit a non-trivial CS invariant
from a Wilson line on the quotient Calabi-Yau.}
\label{slagsmoddedout}
\end{figure}

Now consider the field $\phi|_Q$ restricted to a sLag, $Q$, of $Y_3$. Since
$\Gamma$ acts freely on $Y_3$, we encounter two possibilities for the action
of each generator
$g$ of $\Gamma$ on the sLag $Q\subset Y_3$ (see figure \ref{slagsmoddedout}):
\begin{itemize}
\item $g$ maps $Q$ pointwise to \emph{another} sLag $Q^{\prime}\subset Y_3$ so
that $Q$ and $Q^{\prime}$ are identified in $Y_3/\Gamma$.
In this case, any Wilson line WL$_g$ on $Q$ on the quotient space
$Y_3/\Gamma$ would have to be already present on $Q$ in the covering space
$Y_3$. On $Y_3$, however, the  homotopy loop $\gamma_{g}$
would be contractible so the projection of the Wilson line on $Q$ must
vanish. If this is true for all generators $g$ of $\Gamma$,
it means that all Wilson line operators project to the identity on the sLag
$Q$, and hence they can never
give rise to a non-trivial
 Chern-Simons invariant on $Q$ in the quotient space $Y_3/\Gamma$.
\item If instead $g$ acts freely \emph{within} $Q$, then the corresponding
sLag $Q/\Gamma$ in the quotient space $Y_3/\Gamma$ may acquire a new
 homotopy loop on which the Wilson line on $Y_3/\Gamma$ projects non-trivially.
In this case, there is the possibility to have a non-trivial Chern-Simons
invariant on the sLag $Q/\Gamma$.
\end{itemize}

Having classified a large set of sLags in the CICY as in subsection \ref{involutions}, 
our next task is then to determine how the discrete symmetry $\Gamma$, by which we quotient, acts on them.  Only sLags 
$Q$ that are mapped to themselves by at least one generator $g$ of $\Gamma$, can have non-trivial Wilson lines and hence possible 
Chern-Simons invariants on their quotients $Q/\Gamma$.  As we will now see, this is a model independent question.  Whether or not 
a non-trivial Chern-Simons terms on such a quotient sLag is then really induced, depends also on its topology and the details of 
the Wilson line in $Y_3/\Gamma$ and will be discussed further below.

The discrete symmetry groups used so far in constructing heterotic standard models, and considered in the following, are rotations and
permutations\footnote{See \cite{Braun:2010vc} for a classification of freely acting discrete symmetries on CICYs, including non-Abelian 
ones.  Initial work towards constructing heterotic standard models using non-Abelian discrete symmetries can be found in \cite{Anderson:2009mh},\cite{Anderson:2013xka}.}.  We take 
$\Gamma=\mathbb{Z}_{n+1}\times \mathbb{Z}_{n+1}$, where the first
$\mathbb{Z}_{n+1}$ factor refers to rotations, $R$, and the second to cyclic
permutations, $S$, of the coordinates of $\CP^n$. When we specify how the discrete symmetry group  $\Gamma$ acts on the
coordinates of $\CP^n$, we implicitly fix some or all of the coordinate
freedom of this ambient space.  We give the action of these
symmetries in terms of their respective generators, $g_R$ and $g_S$.
The rotations are generated by
\beq\label{rotation}
g_R: z_a \mapsto \omega^az_a,\qquad a=1,\dots,n+1,
\eeq
where $\omega$ is the primitive $(n+1)$-th root of unity. The generator of
the cyclic permutations acts as
\beq\label{permutation}
g_S: z_i\mapsto z_{i+1}, \qquad i=0,\dots,n, \quad z_0 := z_{n+1} \,.
\eeq
Note that $R$ and $S$ have fixed points on $\CP^n$, but the CICY under
consideration will not contain these fixed points.

\subsubsection{A-type sLags}
We begin by discussing the action of the generators $g\in \Gamma$ on the basic
$A$-type sLags, i.e. the fixed point loci of $\sigma_A$ or, more generally,
$\sigma_A^U$. As these involutions do not mix different ambient $\CP^n$'s,
it is sufficient to restrict our discussion to a single  $\CP^n$ factor.
  
\paragraph{Rotations $R$:}
We first consider the action of the rotations generated by $g_R$ on the
$A$-type sLags. We can treat the basic $A$-type sLag based on the involution
$\sigma_A$ as a special case of the more general case corresponding to
$\sigma_A^U$. The original sLag $Q_{\sigma_A^U}$ is associated with the fixed
point set
$\textrm{Fix}(\sigma_A^U)=\{z\in \CP^n\;|\;  U^{-1}\overline{Uz} = z\}  $.
The rotation $g_R$ maps this to the fixed point set
\begin{equation}
 g_R\fix(\sigma_A^U) = \fix(\sigma_A^{U g_R^{-1}})~.
\end{equation}
Note that due to the projective identification, $z\sim \lambda z$, this is
the same as the original fixed point set  if
\begin{equation}
 g_RU^{-1}\overline{U}g_R= \lambda U^{-1} \overline{U}, \label{gRUm}
\end{equation}
where we used $g_R^{-1}=\overline{g_R}$ and $\lambda$ is a phase factor. Because $g_R^{n+1}=1$, re-iterating this
equation implies $\lambda^{n+1}=1$, i.e.
$\lambda$ is an integer power of the primitive $(n+1)$th root of unity,
$\lambda = \omega^{l}$, $l\in \mathbb{Z}$.
For diagonal $U$, the condition (\ref{gRUm})  becomes
\begin{equation}
 g_R^2= \lambda \mathbf{1},\label{gRm}
\end{equation}
which is only satisfied if $n$, the dimension of the ambient space, equals one.
We therefore see that if $n>1$ the rotational symmetry $g_R$ always maps the
sLag based on $\sigma_{A}^U$  to a different sLag, so that there can be no 
Wilson lines or
Chern-Simons invariant induced by rotational identifications on any $A$-type
sLag.

If on the other hand, the ambient space is $\CP^1$,  the rotational symmetry
is $R\cong \Z_2$ and the  generator $g_R$ automatically satisfies (\ref{gRm})
with $\lambda = 1$. In this case, the generator $g_R$ maps the original sLag
(non-trivially) to itself, and a Chern-Simons invariant might in principle be
induced on any $A$-type sLag by a Wilson line associated with the generator
$g_R$.

In our examples in section \ref{S:explicitmodels}, only the first case with
$n>1$ will occur so that we do not have to worry about  rotational
identifications and  their associated  Wilson lines on $Y_3/\Gamma$.

\paragraph{Cyclic permutations $S$:}
Next we consider the $(n+1)\times (n+1)$ matrices $g_S$ corresponding to the cyclic
permutations (\ref{permutation}).  As they are real, 
the condition for $g_S$ to map a sLag based on the
involution  $\sigma_A^U$ to itself, and hence to induce possible 
Wilson lines and
Chern-Simons
invariants, is not of the form (\ref{gRUm}), but rather:
\begin{equation}
 g_SU^{-1}\overline{U}g_S^{-1}= \lambda U^{-1} \overline{U} \label{gSUm}.
\end{equation}
Let us now give the most general solution of (\ref{gSUm}) for
a diagonal matrix $U=\textrm{diag}(u_1,\ldots,u_{n+1})$ that is assumed to
be a symmetry of the defining polynomial of the CY-space $Y_{3}$. Obviously,
 $U^{-1}\overline{U}=\textrm{diag}(\mu_1,\ldots, \mu_{n+1})$
with $\mu_{i}\equiv \overline{u}_{i}/u_{i}$, and the left hand side of
(\ref{gSUm})  becomes
\begin{equation}
g_SU^{-1}\overline{U}g_S^{-1}=\textrm{diag}(\mu_{2}, \mu_3,\ldots,\mu_{n+1},\mu_1).
\end{equation}
It is then easily seen that the general solution of (\ref{gSUm}) is given by
\begin{equation}
U^{-1}\overline{U}=\mu_{n+1} ~\textrm{diag}(\lambda, \lambda^2,\ldots,\lambda^n,1), \quad \lambda=\omega^l, \quad l\in \mathbb{Z} \label{Ulambda}.
\end{equation}
Any $A$-type sLag on $Y_3$ based on a matrix $U$ that satisfies this equation
for some $l\in \mathbb{Z}$ is then mapped to itself by $g_S$ and 
possibly gives rise to a non-trivial Wilson line and 
Chern-Simons invariant on the
corresponding quotient sLag.

We now show, however, that in many cases (and in particular in all cases we
study in this paper) this apparent multitude of sLags with  potential
Chern-Simons terms actually collapses to just the basic $A$-type  sLag
corresponding to the simple involution $\sigma_A$ when also the rotational
symmetries $R$ are modded out. More precisely, we show that for $nl$ even, any
$A$-type sLag that satisfies (\ref{Ulambda}) is identified with the basic
$A$-type sLag by modding out the rotation $g_R^{nl/2}$.

In order to prove this, one needs to find an integer $k$ such that $g_R^k~\fix(\sigma_A^U) = \fix(\sigma_A)$, i.e.
\begin{equation}
g_R^k U^{-1}\overline{U} g_R^k \propto \bf{1}~.  \label{UgR1}
\end{equation}
Using (\ref{Ulambda}), the left hand side of (\ref{UgR1}) becomes
\begin{eqnarray}
g_R^{k} U^{-1}\overline{U}g_{R}^{k}&=& \mu_{n+1}~\textrm{diag} (\lambda\omega^{2k},\lambda^{2}\omega^{4k},\dots,\lambda^n\omega^{2nk},1)\nonumber\\
&=& \mu_{n+1}~\textrm{diag} (\omega^{l+2k},\omega^{2(l+2k)},\omega^{3(l+2k)},\dots,\omega^{n(l+2k)},1)~,\label{grkUgrk}
\end{eqnarray}
which is proportional to the identity for $2k = -l \mod n+1 = nl \mod n+1$.
This then implies:
\begin{itemize}
\item $\mathbf{n}$  \textbf{even:} Every $A$-type sLag that satisfies
(\ref{Ulambda}) is mapped to the basic $A$-type sLag corresponding to 
$\sigma_A$ by the rotation $g_R^{nl/2}$.
Thus, for $\mathbb{Z}_{\textrm{odd}}$, one only has to check whether this \emph{basic}
$A$-type sLag inherits a Chern-Simons invariant from the Wilson line
associated with the permutation $g_S$.
\item $\mathbf{n}$ \textbf{odd:} In this case, all $A$-type sLags that satisfy
(\ref{Ulambda}) with  $l$ \emph{even} are also identified with the basic
$A$-type sLag upon modding out by $g_R^{-l/2}$ and hence don't have to be
studied separately. On the other hand, the sLags that satisfy (\ref{Ulambda})
with $l$ \emph{odd} are not mapped to the basic $A$-type sLag, but rather the
one corresponding to $\sigma_A^{\sqrt{g_R}}$. 
This is because if we choose $k$ such
that $l + 2k = -1$
 mod $n+1$, we see that eq. \eqref{grkUgrk} implies
\begin{equation}
g_R^k U^{-1}\overline{U}g_{R}^{k} \propto g_R^{-1} = g_{R}^{-1/2}\cdot \overline{g_{R}^{1/2}}~.
\end{equation}
It should be noted that for $n$ odd, $g_{R}^{1/2}$ is 
in general not a symmetry of the polynomial, but still satisfies (\ref{involutioncondition}) because $\sigma_{A}^{\sqrt{g_{R}}}= g_{R}^{-1/2}\circ \sigma_{A} \circ g_{R}^{1/2} = g^{-1}_{R}\circ \sigma_{A}$ and $g_R$ is by assumption a symmetry of the polynomials. Hence $\textrm{Fix}(\sigma_{A}^{\sqrt{g_{R}}})$ is still a sLag, but it is not necessarily homoeomorphic to the basic A-type sLag.

For $n$ odd, we therefore may have possible non-trivial
Chern-Simons invariants on the basic $A$-type sLag and \emph{one} other
$A$-type sLag corresponding to $\sigma_{A}^{\sqrt{g_{R}}}$, which have to be studied separately. In our examples, however,
$n$ is always even and this case does not occur.
\end{itemize}
\paragraph{To summarize:}  If one mods out by the group
$\Gamma= R\times S\cong \mathbb{Z}_{n+1}\times\mathbb{Z}_{n+1}$ of  rotations
\emph{and} cyclic permutations, and if  $(n+1)$ is \emph{odd}, the only
$A$-type sLag one has to check for a possible Chern-Simons invariant is the
basic one based on the simple involution $\sigma_A$, and one only has to consider Wilson lines due to $g_{S}$. 
This will be the case for
all the examples discussed in section \ref{S:explicitmodels}. If $(n+1)$ is
\emph{even}, by contrast, one further $A$-type sLag might carry non-trivial
Chern-Simons invariants on the quotient space $Y_3/\Gamma$ due to modding out
cyclic permutations $S$. For the special case $(n+1)=2$, Chern-Simons
invariants might also occur from modding out certain rotations $R$
(see table \ref{T:tabwlonslags}).

\subsubsection{C-type sLags}
The $C$-type sLags are fixed point sets of involutions (\ref{sigmaC}) or
(\ref{UsigmaC}) that involve the exchange of the coordinates of two
$\CP^n$-factors in an ambient space $\CP^n\times\CP^n$.
This leaves some freedom in defining the action of the symmetries $R$
and $S$ on each factor.  We will consider transformations generated by
$(g_R,g_R^{-1})$ and $(g_S,g_S)$, as these are precisely of the form we
will encounter in our explicit examples in section \ref{S:explicitmodels}.

To begin with, let us recall the fixed point sets of the involutions
$\sigma_C$ and $\sigma_C^{(U_1,U_2)}$:
\bea
\fix(\sigma_C)&=&\{ (z,w)\in\CP^n\times\CP^n \; |\;  \overline{z}=w \}\\
\fix(\sigma_C^{(U_1,U_2)})&=&\{ (z,w)\in \CP^n\times\CP^n \; |\;  U_2^{-1}\overline{U_1z}=w\}, \label{Cfixeq}
\eea
where $U_1$ and $U_2$ are independent elements of $GL(n+1, \mathbb{C})$.
Due to the projective identifications, two sLags associated to
$\sigma_C^{(U_1,U_2)}$ and $\sigma_C^{(U'_1,U'_2)}$ are equivalent whenever
$U_2^{-1}\overline{U}_1=\lambda {U'}_2^{-1}\overline{U'}_1$ for some
$\lambda\in \C$.

We now consider the action of the discrete symmetries $R$ and $S$ on $C$-type
sLags.
\paragraph{Rotations:}
From (\ref{sigmaC}), (\ref{fixsigmaC}) and (\ref{UsigmaC}) one can see that
the generator of the rotation, $(g_R,g_R^{-1})$, acts on the sLag associated to
$\sigma_C^{(U_1,U_2)}$ in the following way:
\beq
(g_R,g_R^{-1}) \fix(\sigma_C^{(U_1,U_2)})= \{ (z,w)\in \CP^n \times \CP^n \; |\;  g_R^{-1} U_2^{-1}\overline{U_1}g_R \overline{z}=w\}.
\eeq
We are interested in the case when this action maps a given sLag non-trivially
to itself. This is the case when
\beq \label{CgRUm}
g_R^{-1} U_2^{-1}\overline{U}_1g_R= \lambda U_2^{-1}\overline{U}_1,\qquad \lambda \in \C.
\eeq
Note that this equation differs from (\ref{gRUm}) in an important way because
the first $g_R$ is inverted. When $U_1$ and $U_2$ are diagonal matrices and
commute with $g_R$, 
 eq. (\ref{CgRUm}) is always satisfied
for \emph{any} $U_1,U_2$. Hence, $(g_R,g_R^{-1})$ acts freely within each $C$-type
sLag associated to $\fix(\sigma_C^{(U_1,U_2)})$. A Wilson line can thus project
non-trivially to any of them, and hence all $C$-type sLags could a priori inherit
a Chern-Simons invariant from a Wilson line associated with modding out a
rotation.

\paragraph{Cyclic permutations:}
The generator, $(g_S,g_S)$, of a cyclic permutation maps the fixed point set
of a $C$-type involution $\sigma_{C}^{(U_1,U_2)} $ to itself
whenever the following equation is satisfied:
\beq \label{CgSUm}
g_S U_2^{-1}\overline{U}_1g_S^{-1}= \lambda U_2^{-1}\overline{U}_1,\qquad \lambda =\omega^l.
\eeq
In contrast to $g_R$, $g_S$ is not diagonal, and hence it does not in general 
commute with $U_2^{-1}\overline{U}_1$. In analogy with the $A$-type involutions,
we have $U_2^{-1}\overline{U}_1=\textrm{diag}(\mu_1,\ldots, \mu_{n+1})$
with $\mu_{i}\equiv \overline{u}^{(2)}_i/u^{(1)}_i$, where $u^{(j)}_i$ is the ith
diagonal element of $U_j$, so that the left hand side of (\ref{CgSUm})  becomes
\begin{equation}
g_SU_2^{-1}\overline{U}_1g_S^{-1}=\textrm{diag}(\mu_{2}, \mu_3,\ldots,\mu_{n+1},\mu_1).
\end{equation}
It is then easily seen that the general solution of (\ref{CgSUm}) is given by
\begin{equation}
U_2^{-1}\overline{U}_1=\mu_{n+1} \textrm{diag}(\lambda, \lambda^{2},\ldots,\lambda^n,1), \quad \lambda=\omega^l \label{CUlambda}.
\end{equation}
Any sLag on $Y_3$ based on matrices $(U_1,U_2)$ that satisfies this equation
for some $l\in \mathbb{Z}$ is thus mapped to itself by $g_S$ and
could possibly give rise to a non-trivial Chern-Simons invariant on the
corresponding quotient sLag.

As we did for $A$-type involutions, we can try to see if we can rotate the sLag
corresponding to such a $\sigma_C^{(U_1,U_2)}$ to the basic one. 
However, this is not
possible here, since, as seen above, any rotation
$(g^m_R,g_R^{-m})$ ($\, \forall m \in \Z_{n+1} $) only maps a $C$-type sLag to
itself.

\paragraph{To summarize:} Wilson lines associated with permutations $S$ and
rotations $R$ may project non-trivially to the basic $C$-type sLag, which could
thus inherit a non-trivial Chern-Simons invariant from both these Wilson lines.
The more general $C$-type sLags associated to $\sigma_C^{(U_1,U_2)}$, on the other
hand, are likewise sensitive to any Wilson lines associated to $R$, but carry Wilson line projections corresponding  to permutations $S$ only
when  (\ref{CUlambda}) is satisfied.  Thus these general $C$-type sLags have to
be checked for corresponding Chern-Simons invariants as well (see table
\ref{T:tabwlonslags}).

\begin{table}[ht]    
\centering
\begin{tabular}{| p{2.8cm} || l | l | l | l |}
\hline
 $\Gamma= \Z_{n+1} \times \Z_{n+1}$ & $\fix(\sigma_A)$ & $\fix(\sigma^U_A)$ & $\fix(\sigma_C)$  & $\fix(\sigma^{(U_1,U_2)}_C)$  \\ \hline \hline
 $n=1$ & $g_R$, $g_S$ &  $g_R$, $g_S \blacklozenge$ & $g_R$, $g_S$ & $g_R$, $g_S^{\clubsuit}$ \\ \hline
  $n$ even & $g_S$ &  $g_S \lozenge$  & $g_R$, $g_S$ & $g_R$, $g_S^{\clubsuit}$ \\ \hline
  $n>1$ odd & $g_S$ &  $g_S \blacklozenge$  & $g_R$, $g_S$ & $g_R$, $g_S^{\clubsuit}$\\
   \hline
   \end{tabular}
\caption{\label{T:tabwlonslags} In the table we summarize the cases encountered for the action of the
generators $g_R$ and $g_S$ of the symmetry group $\Gamma=R \times S\cong
\Z_{n+1} \times \Z_{n+1}$ on the $A$- and $C$-type sLags . In the first row we
label the sLags associated to their involutions. The entries indicate which
generators map the sLags non-trivially into themselves and hence could potentially induce non-trivial Chern-Simons invariants. The symbol $\lozenge$
means that the corresponding sLag is mapped into $\fix(\sigma_A)$ by the
action of $R$ if \eqref{UgR1} is satisfied so that one does not have to study
it separately for the Wilson lines of $g_{S}$. 
The symbol $\blacklozenge$ indicates that the corresponding sLag
is either mapped to $\fix(\sigma_A)$ (and hence does not have to be studied separately)
or to $\fix(\sigma_{A}^{\sqrt{g_{R}}})$ by the action of $R$ (if \eqref{UgR1} is satisfied).
Which of these two possibilities is realized depends on whether $l$ in \eqref{UgR1} is even or odd, respectively. The superscript $ ^{\clubsuit}$, finally, means that the generator $g_S$ maps
the sLag into itself only if \eqref{CUlambda} is satisfied. }
\end{table}

\subsection{Chern-Simons invariants on Seifert fibered 3-manifolds}
\label{CSonSeifert}
In the previous subsections, we have provided a classification of particular
3D submanifolds,
sLags, that can be explicitly constructed in CICYs.  We have also considered
how Wilson lines in a 
CICY project onto these sLags.  The next step in computing the flux
superpotential due to Wilson lines
is to compute the Chern-Simons invariants on the sLags on which the
Wilson
lines project non-trivially.  Therefore, in this subsection, we will give some
general mathematical
results relevant to computing Chern-Simons invariants on
a large class of
closed,
compact, orientable 3D (sub)manifolds.
As we will see, a class of 3D manifolds very widely encountered are
so-called
Seifert fibred manifolds, or compositions thereof.

  We will apply the results presented here to treat our
explicit examples in the next section, and indeed expect them
to be useful more generally.  This section is a somewhat technical summary of
the mathematical
literature, and the reader may wish to skip it on the first read.

\paragraph{Decomposition theorems}
We begin by discussing two important ways to simplify the description of a 3-manifold, by decomposing it into more basic pieces \cite{hatcher}. 

The first is called a \emph{prime decomposition}; every compact orientable 3-manifold $M$ has a unique decomposition along 2-spheres as a connected sum\footnote{The connected sum of two 3-manifolds is formed by deleting a 3-ball from each, and gluing together the resulting boundary 2-spheres.} $M = P_1 \sharp \dots \sharp P_n$, where each $P_i$ is a prime manifold (i.e., the only way that $P_i$ splits as a connected sum is the trivial one $P_i=P_i \sharp S^3$).  Note that a prime manifold is either irreducible (every 2-sphere bounds a ball) or diffeomorphic to $S^2 \times S^1$.  

The second is called a \emph{torus decomposition}; every irreducible compact orientable 3-manifold $M$ can be decomposed by cutting along incompressible 2-tori $T_i$ (i.e., a torus $T_i$ such that the induced map $\pi_1(T_i) \rightarrow \pi_1(M)$ is injective), to give the union $M=X_1 \cup \dots \cup X_n$, where each $X_i$ is either Seifert fibered or atoroidal (i.e., every incompressible torus in $X_i$ is isotopic to a torus component of $\partial X_i$).  Note that atoroidal 3-manifolds are hyperbolic.

The sLags we encounter in our concrete CICY examples indeed simply turn out to be Seifert fibered manifolds, or can be decomposed into Seifert fibered manifolds using a torus decomposition. 

\begin{figure}[h]
\begin{center}
\resizebox{10cm}{!}{\input{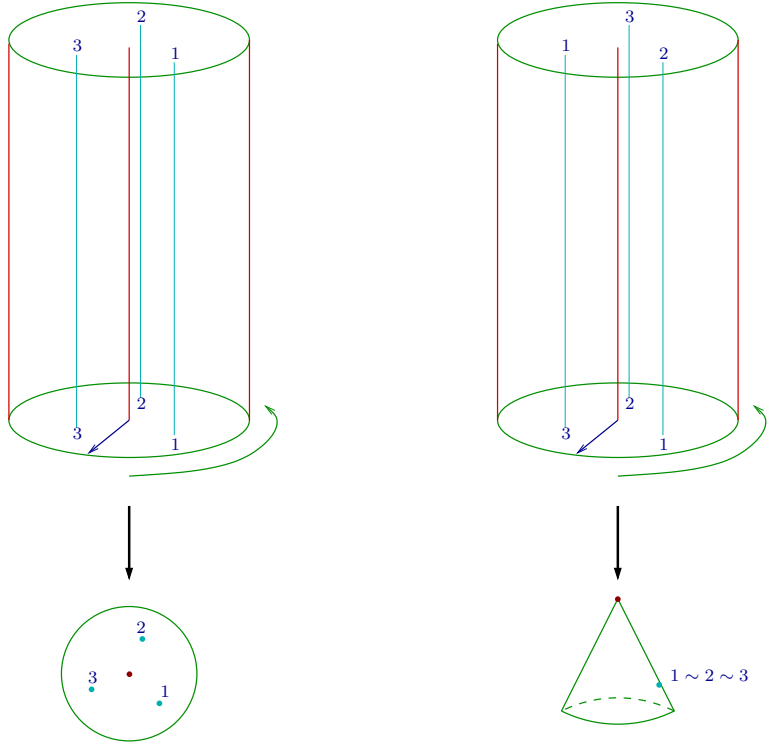}}
\caption{Fibered solid tori in a Seifert fibration.  On the left we show an ordinary solid torus, and on the right a fibered solid torus.  
They are $D \times I$ with $D$ being the unit disk in $\mathbb{C}$ and the ends of the interval $I$ identified, and fibered by the 
intervals $\left\{x\right\}  \times I$ with $x\in D$. 
Defining the homeomorphism $\rho: D\rightarrow D$ by $\rho(x) = x e^{2\pi\i q/p}$, we construct the fibered solid torus by identifying $(x,0)$ 
with $(\rho(x),1)$.  The integers $p,q$ are co-prime, and are chosen to satisfy $0\leq q<p$. The ordinary solid torus has $(p,q) = (1,0)$ and the 
fibered solid torus depicted has $(p,q)=(3,1)$.  The central fiber of the fibered solid torus $\left\{0\right\}  \times I$ is called the exceptional 
fiber.  It covers the interval $I$, and intersects the disk $D$, once.  The other fibers are regular fibers.  They cover the interval -- and intersect 
the disk $D$ -- a multiple $p$ times before closing.  Taking the quotient space of a Seifert fibered manifold by identifying all circular fibers to a 
point results in a 2-dimensional orbifold $B$, with orbifold points at the location of the exceptional fibers, as illustrated at the bottom of the figure.}
\label{F:regularsingularfiber}
\end{center}
\end{figure}

\paragraph{Seifert fibered manifolds}
Seifert fibered manifolds are among the best understood 3D manifolds, and their Chern-Simons invariants can be explicitly calculated 
using the results of \cite{Auckly,Nishi:1998}.   
Let us start with a definition of Seifert
fibered manifolds (see e.g. \cite{neumann_lectures, hatcher, brin,montesinos1987} for some
lectures on these spaces):
A Seifert fibered manifold, $Q_{Sf}$, is a 3D manifold that is a
union of pairwise disjoint circles (the fibers) such that
the neighborhood of each circle fiber is diffeomorphic to a, possibly fibered, solid torus.\footnote{In case $Q_{Sf}$ has 
boundaries, the boundary fibers are located on the boundary of a suitable fibered solid torus.}  Equivalently, a Seifert fibered 
manifold can be described as an $S^1$ fibration over a 2-dimensional orbifold base called the orbit surface.  The fibered solid torus 
and orbifold surface and the relation between them are explained in figure \ref{F:regularsingularfiber}.  

A Seifert fibration is characterized by a so-called Seifert invariant, 
which is the collection of relevant topological data,
\beq
Q_{Sf}=\left\{O,o,g;b,(\alpha_1,\beta_1),\dots,(\alpha_s,\beta_s)\right\}~. \label{Sfinv}
\eeq
Here, the symbol $O$ denotes that the Seifert fibered manifold is orientable
and the symbol $o$ denotes that the orbit surface is orientable\footnote{We only
consider orientable Seifert fibered manifolds and orbit surfaces in this paper, but this restriction can easily be lifted.}, $g$
is the genus of the orbit surface, $b$ is called the section obstruction of the Seifert fibration\footnote{More precisely the section 
obstruction refers to the circle bundle with no exceptional fibers, which is obtained by \emph{drilling out} the fibered solid tori of the Seifert 
fibered manifold and \emph{filling in} with standard solid tori; the resulting smooth fibration has global section iff $b=0$. We refer to
 \cite{brin,montesinos1987} for more details.}
which vanishes for manifolds with non-empty boundary, 
$s$ is
the number of exceptional fibers, i.e. the number of orbifold points in the
base, and the pairs $(\alpha_j,\beta_j)$ (with $j=1,\dots,s$) describe the exceptional fibers.  For each exceptional fiber, the invariant 
$(\alpha_j,\beta_j)$ is given in terms of the invariant $(p_j,q_j)$, which describes the associated fibered solid torus as in 
figure \ref{F:regularsingularfiber}, by $\alpha_j = p_j$ and
\beq
0 < \beta_j < \alpha_j, \quad \beta_j q_j \equiv 1 \mod \alpha_j \,.
\eeq
Note that one and the same Seifert fibered manifold might be describable in
terms of different Seifert invariants
in case it admits several ways of splitting it into base and fibers.

Finally, in order to describe Wilson lines and Chern-Simons invariants on Seifert fibered manifolds, one needs to know their fundamental groups.  
A presentation of the fundamental group of a Seifert fibration can be read off directly from the Seifert invariant, with the generators and relations 
given by \cite{brin}:
\bea
\pi_1(Q_{Sf}) =&& \langle h,a_1,b_1, \dots, a_{g},b_g, \, c_0, c_1, \dots, c_s, \, d_1,\dots d_m, \; \vline h\text{ is central}\nonumber \\ && c_0h^b = c_j^{\alpha_j} h^{\beta_j} = \prod [a_i,b_i]\prod c_j\prod d_k =1 \rangle\,, \label{Sfpi1}
\eea
where $i=1,\dots,g$, $j=1,\dots,s$ and $k=1,\dots, m$, and $m$ is the number of boundary components of the 3-manifold. As a simple illustration, 
consider e.g. the 3-torus as a trivial $S^{1}$-bundle over the orbit surface $T^{2}$, so that $g=1$, $b=0$ and  $s=0$, and hence eq. (\ref{Sfpi1}) 
gives $c_{0}=1$ and three commuting non-trivial generators $h, a_1, b_1$, i.e. the expected result $\pi_{1}(T^{3})=\mathbb{Z}^{3}$.

\paragraph{Chern-Simons invariants on Seifert fibered manifolds and their compositions}
We now summarize some known results for Chern-Simons invariants on closed
Seifert fibered manifolds, and closed manifolds that decompose into Seifert fibered manifolds with boundary under a torus decomposition. 
 The Chern-Simons invariant for all flat $SU(2)$
connections on all closed
Seifert fibered spaces was computed in \cite{Auckly}.  The Chern-Simons
invariant for a general class of flat $SU(N)$ bundles on any closed Seifert
fibered 3-manifold was computed in \cite{Nishi:1998}.
These results are stated in terms of irreducible and reducible flat connections
(a reducible flat connection is one for which the subgroup $H$ commuting with
the image of the
homomorphism $\rho:\; \pi_1(Q_{sf}) \rightarrow G$ has continuous parameters, otherwise
it is irreducible\footnote{A sufficient, but not necessary, condition for a connection $\rho:\; \pi_1(Q_{Sf}) \rightarrow G$ to be reducible is 
that $\rho(h)$ lies outside the center of $G$.  In these cases, all elements of $\pi_1(Q_{Sf})$ must map to the Cartan subalgebra, and $H$ is at 
least $U(1)^r$ with $r$ the rank of $G$.}).  Notice that the Wilson lines of interest to us are always
reducible flat connections,
because $H$ should always contain the gauge group of the Standard Model.
Moreover, our Wilson lines always lie in a maximal torus of the gauge group
$G$.  The Chern-Simons invariant for Abelian reducible $SU(N)$ connections with $\rho:\; \pi_1(Q_{sf}) \rightarrow SU(N)$ given by $\rho(h) =
\exp{2\pi\i Y}$, $\rho(c_j) = 1$, on Seifert fibered 3-manifolds without
boundaries is\footnote{This result follows from the expression given above Lemma 3.3 in \cite{Nishi:1998}.  Indeed, we need to relax the condition 
applied in Lemma 3.3 that $\rho(h)$ be a scalar matrix, as the Wilson lines encountered are typically not scalar matrices.} \cite{Nishi:1998}
\beq
CS(A,Q_{Sf}) = \frac12 b \, \tr Y^2 + \frac12 \sum_{j=1}^s \beta_j\, \delta_j \,\tr Y^2 \mod \Z \label{CSSf}
\eeq
where $Y$ is in the Lie algebra of a maximal torus of $SU(N)$ and $\delta_j \in \Z$ is such that
$\alpha_j \delta_j - \beta_j \gamma_j = 1$ for some integers $\gamma_j$. It is immediate that for the 3-torus with $b=0=s$ this Chern-Simons invariant 
is zero (modulo integers), as we will use later.

In our examples, we will also encounter sLags that are not Seifert fibered
manifolds, but reduce to Seifert fibered manifolds with boundary
under a  torus decomposition.
For such more general manifolds, we may use the results of \cite{KirkKlassen2},
where it was shown how to compute Chern-Simons invariants on 3-manifolds
decomposed along tori\footnote{Ref. \cite{KirkKlassen2} also considers cases
when some components of the torus decomposition are not Seifert fibered but
hyperbolic manifolds.}.  Indeed,
for a 3-manifold $M$ that decomposes into a union of Seifert fibered spaces,
$X_i$, the Chern-Simons invariant on $M$ may be obtained by first computing
the Chern-Simons invariants
on the pieces $X_i$, and then computing the effect of gluing the pieces
together.  Some extra care is required because Chern-Simons invariants on
manifolds with boundary are not gauge invariant, even up to integers.

For example, consider $M$ a closed 3-manifold decomposed along a torus $T$ as
$M=X_1 \cup_T X_2$, and an $SU(2)$ flat connection over it.  The toroidal
boundaries $\partial X_i =T_i$
have fundamental group $\pi_1(T_i) = \left\langle\mu_i, \lambda_i\right\rangle$.
The gluing together of $X_1$ and $X_2$ along their boundaries is described by a
map between these generators:
$\mu_1 \rightarrow p  \mu_2 + q  \lambda_2, \; \lambda_1 \rightarrow r  \mu_2 + s  \lambda_2$, with $ps-qr=1$.  Meanwhile, the restriction of the Wilson lines on $X_i$,
$\rho:\; \pi_1(X_i) \rightarrow SU(2)$, to $T_i$ is given by:
\bea
\rho(\mu_i) = \left( \begin{array}{cc} e^{2\pi\i a_i} & 0 \\ 0 & e^{-2\pi\i a_i}\end{array} \right) \, \qquad \rho(\lambda_i) = \left( \begin{array}{cc} e^{2\pi\i b_i} & 0 \\ 0 & e^{-2\pi\i b_i}\end{array} \right) \,.
\eea
We then define equivalence classes of Chern-Simons invariants on each $X_i$:
\beq
\left[\left\{a_i, b_i; \; e^{2\pi\i CS(A,X_i)} \right\}\right]\,,
\eeq
where the square brackets indicate the orbit of $SU(2)$, with the equivalence relation:
\beq
\left\{a_i, b_i; e^{2\pi\i CS(A,X_i)} \right\} = \left\{a_i+m, b_i+n;  e^{2\pi\i (m  b_i - n a_i)}e^{2\pi\i CS(A,X_i)} \right\}
\eeq
for $m,n \in \Z$.  Finally, the Chern-Simons invariant on $M$ is defined as the inner product:
\beq
CS(A,M) = \langle CS(A,X_1), CS(A,X_2) \rangle \,,
\eeq
which is simply given by the sum $CS(A,X_1) + CS(A,X_2)$ after choosing gauge fixings that are compatible  with the gluing map, 
$a_1 = p  a_2 + q  b_2, \; b_1 = r  a_2 + s  b_2$.

\subsection{The superpotential from Chern-Simons invariants}
Before considering some explicit examples, let us here outline the full
procedure for computing the superpotential due to Chern-Simons fluxes from
Wilson lines.

\begin{enumerate}

\item Identify sLags in a given quotient CICY via its isometric
anti-holomorphic involutions of type $A$ and $C$.
If the discrete group is $\Gamma = R\times S$ with $R$ and $S$
cyclic groups of odd order, then only the basic $A$-type sLag could inherit a Wilson line associated only with $S$. For the
$C$-type sLags, on the other hand, all can inherit Wilson lines associated with
$R$, and sometimes also associated with $S$. The case of even order cyclic
groups does not occur in our examples but a complete discussion on which sLags
are relevant or not is given in section \ref{S:WLonsLags}. 

\item Compute the intersection matrix for sLags on the quotient CICY.  If the
rank of the intersection matrix equals the dimension of the third homology
group, then the sLags
constitute a basis for the 3-cycles in the quotient CICY.  In this case, we
can write the 3-cycle, $\Lambda$, Poincar\'e dual to the holomorphic 3-form,
as
\beq
\Lambda = \frac{4}{\alpha^{'}} \sum_K c^K Q_K~,
\eeq
in homology, where $Q_K$ are the sLags, satisfying the specialness condition
with various calibration angles (so $\Lambda$ is in general not sLag), and $c^K$ are
constant
coefficients that depend on the complex structure moduli.
Therefore, the background superpotential is given
by\footnote{$\int_C \phi = \int_{C'} \phi$ for $C$ and $C'$ in the
same homology class and $\phi$ closed.  In the vacuum,
$\d\omega_{3\text{Y}}=0$.},
\beq
W = -\frac{\alpha^{'}}{4} \int_{Y_3} \omega_{3Y} \w \Omega = -\frac{\alpha^{'}}{4} \int_{\Lambda} \omega_{3Y} = -\sum_Kc_K \int_{Q_K} \omega_{3\text{Y}} = -\sum_Kc_K ~CS(A, Q_K) \,,
\eeq

\item  
Study the topology of the $A$-type and $C$-type sLags of the
modded out CICY.  For the
sLags on which the Wilson lines project, one then has to compute the Chern-Simons invariants, and finally write down the
explicit superpotential.  For example, suppose the Chern-Simons invariant is
non-trivial only on the basic $A$-type sLag, and that
the $A$-type sLags are Lens spaces $L(p,1)$  (we will see below that this is the
case for the $\Z_5\times \Z_5$ quotient of the Fermat quintic).  Then, using
\eqref{CSLens}, we have for the superpotential
in the vacuum,
\beq
W = -c~ CS(A,Q_{\sigma_A}) =c \left(\sum_i \frac{k_i^2}{2p} \mod \Z \right)~.
\eeq
\end{enumerate}

Should we wish $W=0$ in the vacuum, due to any of the reasons mentioned in section
\ref{S:3formflux}, we require the Chern-Simons flux on $Q_{\sigma_A}$ to be vanishing
(assuming a non-vanishing value $c$), and
this provides a constraint on the Wilson lines that can be introduced in any
explicit model. In the example above, the necessary and sufficient condition
is that the Wilson lines satisfy
\beq
\sum_i \f{k_i^2}{2p} = 0 \mod \Z\,. \label{constrainteq}
\eeq
The same result would be a necessary condition for setting $H=0$, even if the
third homology group were not spanned by sLags.

Note that although the Chern-Simons invariants are (fractionally) quantized,
the coefficients $c_K$ may take on more general values. In principle, the vacuum expectation value
of $W$ might thus be accidentally small leading to additional suppression of
the gravitino mass in the scenario discussed in \cite{Gukov:2003cy}. 
It is not clear whether this is actually possible; it was argued in \cite{Cicoli:2013rwa} that moduli stabilization
from Chern-Simons flux and gaugino condensation generically leads to
high-scale supersymmetry breaking.

\section{Concrete examples}
\label{S:explicitmodels}
In this section we will apply our strategy to compute the Chern-Simons flux
superpotential in explicit compactifications.  Several of the steps are model
dependent, in particular the computations of the sLag intersection matrix and
the sLag Chern-Simons invariants.  We therefore begin this programme by
treating two concrete examples.  Although not realistic, the four generation
quintic quotient  provides a simple first example to illustrate our arguments.
We will then progress to the three generation split-bicubic quotient, which
has a potentially realistic particle spectrum, corresponding to the MSSM,
a hidden sector and moduli.

\subsection{The four generation quintic quotient}

The Fermat quintic, $X^{1,101}$, is defined by the following hypersurface in
$\CP^4$:
\beq\label{fermatquintic}
X^{1,101} = \left\{ z\in \CP^4\ \bigg|\ \sum_{i=1}^5  z_i^5 = 0\right\}.
\eeq
The notation $X^{1,101}$ refers to the two non-trivial Hodge numbers
$(h^{1,1},h^{1,2}) = (1,101)$.
The quintic has two freely acting order five symmetries, each isomorphic to
$\Z_5$, generated respectively by:
\bea
g_R: (z_1, z_2, z_3, z_4, z_5) &\rightarrow& (\omega z_1, \omega^2 z_2, \omega^3 z_3, \omega^4 z_4, z_5) \nn\\
g_S: (z_1, z_2, z_3, z_4, z_5) &\rightarrow& (z_5, z_1, z_2, z_3, z_4)
\eea
with $\omega=\e^{2\pi\i/5}$. These are precisely the symmetry groups $R$ and $S$
discussed in section \ref{S:WLonsLags}.

A four-generation model \cite{Green:1987mn} can be constructed by compactifying
on the quintic quotiented by $\Gamma=R\times S$, to give non-trivial
fundamental group $\pi_1(Y_3) = R\times S \cong \Z_5 \times \Z_5$.  The choice
of vector bundle corresponding to the standard embedding breaks the
$E_8 \times E_8$ gauge group to $E_6 \times E_8$.  Depending on the choice of
Wilson lines, the $E_6$ is broken further to some extension of the Standard
Model gauge group with chiral matter representations.  We will take just one
of the two possible Wilson lines, associated with either $R$ or $S$, to be
non-trivial.  Using $E_6$'s maximal subgroup
$SU(3)_c \times SU(3)_L \times SU(3)_R$, we can write the Wilson line as the
$27\times 27$-matrix,
\beq
{\rm WL}_\gamma = (\mathbf{1}_3)_c \otimes \textrm{diag}(\alpha,\alpha,\alpha^{-2})_L\otimes \textrm{diag}(\beta, \rho,\delta)_R\,, \label{quinticWL}
\eeq
with $\alpha^5=\beta^5=\rho^5=\delta^5=1$ and $\beta\rho\delta=1$, which is the
most general $\text{WL}_\gamma$ that commutes with the SM gauge group.  E.g. for
$\beta=\rho=\alpha$ and $\delta=\alpha^{-2}$, the unbroken gauge group is
$SU(3)_c\times SU(2)_L\times SU(2)_R \times U(1)^2$.  The Hodge numbers of the
quintic quotient, $X^{1,5}$, are $(h^{1,1}, h^{1,2})=(1,5)$.

The Fermat quintic has a number of isometric anti-holomorphic involutions,
whose actions are not free, and whose fixed points correspond to special
Lagrangian submanifolds \cite{Brunner:1999jqpa}.  The involution
$\sigma_A:z_i \mapsto \bar{z}_i$ has as fixed points the real quintic
\beq\label{basicAtypeslaginquintic}
Q_{\sigma_A} = \fix(\sigma_A)\cap X^{1,101} = \RP^4 \cap X^{1,101} = \left\{ x\in \RP^4\ \bigg|\ \sum_{i=1}^5 x_i^5 = 0\right\}~.
\eeq
One of the coordinates, say $x_5$, can always be expressed uniquely in terms of
the other coordinates which are completely unrestricted but just subject to the
projective rescaling. This means that $Q_{\sigma_A}$ is topologically
$\RP^3 \cong S^3/\Z_2$ (note that this is a Lens space and hence also a
Seifert fibered manifold). As discussed above we can construct many more $A$-type
involutions by considering $\sigma_A^U = M\circ \sigma_A$ where $M=U^{-1}\overline{U}$,
and $U$ is a symmetry of the defining polynomial of the quintic. Taking only
diagonal matrices $U$, we get $5^4=625$ non-trivial and distinct involutions of
this type. The fixed point loci of these involutions are given by
\beq
Q_{\sigma_A^U} = \fix(\sigma_A^U)\cap X^{1,101} = U^{-1}(Q_{\sigma_A})\cong \RP^3.
\eeq
By computing the intersection matrix, one can show that only 204 of the 625
sLags $Q_{\sigma_A^U}$ are distinct in homology, and that they span the homology
group of the quintic $X^{1,101}$ \cite{Brunner:1999jqpa} (see appendix
\ref{S:appendix}).

Now let us consider the four-generation quintic quotient $X^{1,5}$.  The number of distinct $A$-type sLags on the quotient
$X^{1,5}$ can be computed to be 129 (see appendix \ref{S:appendix}), and the
rank of the $129 \times 129$ dimensional intersection matrix is reduced to 12.
This matches the dimension of the third homology group for the quintic
quotient, so that the sLags continue to provide a basis for the 3-cycles, as
expected.  We have
seen in subsection \ref{S:WLonsLags} that the only $A$-type sLag one has to
check for a non-trivial Wilson line, is the basic one
\eqref{basicAtypeslaginquintic}. Moreover, this basic $A$-type sLag can at
most inherit Wilson lines, and hence Chern-Simons invariants, from the
permutation group $S$.  

We can immediately write down the full Chern-Simons flux superpotential.  Choosing to embed the Wilson line only in $R$, all the Chern-Simons
invariants are trivial, and therefore, the superpotential is also trivial.
Embedding instead the Wilson line in $S$, the only non-trivial Chern-Simons
invariant is on the sLag $Q_{\sigma_A}$, which on the quotient is the Lens space
$\RP^3/\Z_5=S^3/\Z_{10}$.  Writing
$\alpha=e^{2\pi\i 2k_1/10}$, $\beta=e^{2\pi\i 2k_2/10}$, $\rho=e^{2\pi\i 2k_3/10}$ and
$\delta=e^{2\pi\i 2k_4/10}$ ($k_{1,2,3,4}=0,\dots, 4$) in \eqref{quinticWL}, and
using \eqref{CSLens}, the Chern-Simons invariant is immediately given by
\beq
CS(A,Q_{\sigma_A}) = -\frac{9}{5} \left(6 k_1^2 + k_2^2 + k_3^2 + k_4^2\right) \mod \Z\,,
\eeq
which reduces to $CS(A,Q_{\sigma_A})=-\frac{108}{5} k_1^2 \mod \Z$ for the
$SU(3)_c\times SU(2)_L\times SU(2)_R \times U(1)^2$ model.  The full
superpotential from the visible sector Wilson lines in the vacuum is then
simply:
\beq
W=c \left(\frac{108}{5} k_1^2 \mod \Z \right)=c \left(\frac{3}{5} k_1^2 \mod \Z \right)
\eeq
for $c$ a (possibly) non-vanishing constant, depending on the choice of
complex structure. The $\mod\Z$ can be interpreted as a possible integer
$H$-flux contribution.  There may also be non-trivial contributions from
hidden sector Wilson lines, which could e.g. be chosen to ensure two or more
condensing gauge sectors to help stabilize moduli.  Of course, the hidden
Wilson lines project in the same way as the visible ones on each sLag, and
they only differ in their explicit values.

\subsection{The three generation split-bicubic quotient}
\label{SS:splitbicubic}
We now turn to a potentially realistic compactification, based on a quotient of
the split-bicubic CY threefold \cite{Ovrut:2002jk, Braun:2004xv}.  After introducing the CICY and its quotient we will follow the same procedure as above, which is here somewhat more involved.  We identify the $A$-type and $C$-type sLags, and study their topology, particularly in the quotient CICY.  Then we can compute the relevant Chern-Simons invariants by using the torus decomposition into Seifert fibered manifolds, discussed in section \ref{CSonSeifert}.  Finally, we compute the intersection matrix for the sLags and show that we can generate the full third homology group.  In this way, we obtain the full Chern-Simons flux superpotential.

\paragraph{The split-bicubic CICY} It
will be useful to have several pictures of the split-bicubic in mind.  The
first is as a Schoen manifold, which is a fiber product of two rational
elliptic surfaces, $B$ and $B'$, with a common base $\CP^1$,
\beq
X^{19,19}=B \times_{\CP^1} B'=\{ (b,b') \in B \times B'\; |\; \beta(b)=\beta'(b')\},
\eeq
where
\beq
\beta: B \rightarrow \CP^1, \qquad \beta': B' \rightarrow \CP^1,
\eeq
are the projections of $B$ and $B'$ on the common $\CP^1$-base. This can be
represented by the following pull back diagram
\begin{displaymath}
\xymatrix{
& \qquad {X^{19,19}} \ar[dl]_\pi \ar[dr]^{\pi'} & \\
B \ar[dr]_\beta & & B' \ar[dl]^{\beta'} \\
& \CP^1 &
}
\end{displaymath}
so that the CY admits a fibration over $\CP^1$ with generic fiber the product of two elliptic curves.  
The rational elliptic surfaces $B, B'$ are known as $dP_9$, due to their
similarity to the del Pezzo surfaces. Indeed, $dP_9$ is a blow up\footnote{A blow up of an $n$-dimensional complex manifold, $M$, at 
$m$ points to $\CP^1$ is diffeomorphic to the connected sum $M \sharp_m \overline{\CP^2}$, where $\overline{\CP^2}$ has opposite orientation 
to $M$ \cite{McDuffSalamon}.  So $dP_9$ may also be considered as the connected sum $\CP^2\sharp_9 \overline{\CP^2}$.} of $\CP^2$ at nine
points to $\CP^1$ and may be represented by the configuration matrix
\beq
\left[\begin{array}{c|c}
\CP^1 & 1  \\
\CP^2 & 3 \\
\end{array}\right] \,.
\eeq
In other words, it can be written as the hypersurface
\beq
B=\left\{\left(t,\zeta\right)\in\CP^1 \times \CP^2 \;\vline \;t_1 f(\zeta) - t_2 g(\zeta) = 0 \right\},
\eeq
where $t_a$ ($a=1,2$) are homogeneous coordinates of $\CP^1$, $\zeta_j$
($j=1,2,3$) are homogeneous coordinates of $\CP^2$, and $f(\zeta)$ and
$g(\zeta)$ are cubic polynomials.  The equation
$t_1 f(\zeta) - t_2 g(\zeta) = 0 $ can be solved uniquely for $t_a$ in terms of
$\zeta_j$, except for those nine points of $\CP^2$ where $f(\zeta)=0=g(\zeta)$.
At those nine points of $\CP^2$ the $t_a$ are unrestricted and hence
parameterize an entire $\CP^1$.

As there is a similar description for $B'$, the elliptically fibered Calabi-Yau
can also be described as a CICY with the configuration matrix:
\beq\label{splitbicubicmatrix}
\left[\begin{array}{c|cccc}
\CP^1 & 1 & 1 \\
\CP^2 & 3 & 0\\
\CP^2 & 0 & 3
\end{array}\right].
\eeq
In other words,
\bea
X^{19,19} = \left\{\left(t,\zeta,\eta\right)\in\CP^1 \times \CP^2 \times \CP^2 \;\vline \;P_1(t,\zeta) = P_2(t,\eta) = 0 \right\}~,
\eea
where
\bea
P_1(t,\zeta) &=& t_1 f(\zeta) - t_2 g(\zeta), \nonumber \\
P_2(t,\eta) &=& t_1 \hat{g}(\eta) - t_2 \hat{f}(\eta), \label{splitbicubicCICY}
\eea
$\eta_j$ ($j=1,2,3$) are homogeneous coordinates for the second $\CP^2$ factor,
and $f,g,\hat{f},\hat{g}$ are cubic polynomials. When specifying the
polynomials, we have 19 degrees of freedom as the Hodge number $h^{1,2} = 19$
indicates. Here we will make the same choice as in \cite{Candelas:2007ac},
\bea
&&f(\zeta) = \zeta_1^3+\zeta_2^3+\zeta_3^3 - a\,\zeta_1 \zeta_2 \zeta_3,\qquad g(\zeta) = -  c\, \zeta_1 \zeta_2 \zeta_3~, \nonumber\\
&&\hat{g}(\eta) = c\,  \eta_1 \eta_2 \eta_3,\qquad \hat{f}(\eta) = -\eta_1^3 -\eta_2^3 -\eta_3^3 + b\,\eta_1 \eta_2 \eta_3\,. \label{splitbicandelas} 
\eea
This turns out to be the most general choice of polynomials for which the
split-bicubic has a freely acting discrete symmetry $\Gamma=R\times S$ with
$R,S$ both isomorphic to $\Z_3$, with the following
generators\footnote{Another, equivalent, choice is made in
\cite{Braun:2004xv,Braun:2007sn}.} \cite{Candelas:2007ac}:
\bea
&&g_R: \quad \zeta_j \rightarrow \omega^j \zeta_j, \quad \eta_j \rightarrow \omega^{-j} \eta_j, \quad t_a \rightarrow t_a \,, \nonumber \\
&&g_S: \quad \zeta_j \rightarrow \zeta_{j+1}, \quad \eta_j \rightarrow \eta_{j+1}, \quad t_a \rightarrow t_a \,,
\eea
where $\omega=e^{2\pi\i/3}$. The Hodge numbers of the quotient split-bicubic,
$X^{3,3} = X^{19,19}/\Gamma$, are $(h^{1,1},h^{2,1})=(3,3)$. The coefficients
$a,b,c$  in \eqref{splitbicandelas} correspond,
roughly speaking, to the three complex structure moduli of $X^{3,3}$. In order
to analyze the equations explicitly, we will take $a=b=0$ and leave $c=1$.
The polynomials then satisfy
$f =- \hat{f}$, $g = -\hat{g}$ and
\bea
P_1(t,\zeta) &=& t_1 f(\zeta) - t_2 g(\zeta) = t_1\left(\zeta_1^3+\zeta_2^3+\zeta_3^3\right) +  t_2 \zeta_1 \zeta_2 \zeta_3,  \nonumber \\
P_2(t,\eta) &=& t_2 f(\eta) - t_1 g(\eta) = t_2\left(\eta_1^3+\eta_2^3+\eta_3^3\right) +  t_1 \eta_1 \eta_2 \eta_3.
\eea
Note that since $\dd P_1\w \dd P_2$ does not vanish in this case, the
resulting manifold is diffeomorphic to all smooth split-bicubic CICYs. Putting
all three parameters to zero would also be an attractive choice, but corresponds to a
singular limit of $X^{3,3}$.  A heterotic MSSM with no exotics (beyond hidden
sectors and moduli) can be obtained from a compactification on $X^{3,3}$. To
this end, one introduces an $SU(4)$ holomorphic stable vector bundle, and the
following Wilson lines, which embed the $\Z_3 \times \Z_3$ fundamental group
into the $SO(10)$ GUT gauge group using the $\bf{10}$ representation of
$SO(10)$ \cite{Braun:2004xv,Braun:2005nv,Braun:2013wr}:
\bea
\text{WL}_{\gamma_1} = \left(\begin{array}{cc}
e^{4\pi\i/3} \mathbf{1}_{5} &   \\
 & e^{2\pi\i/3}\mathbf{1}_5
\end{array}\right)
\quad
\textrm{and}
\quad
\text{WL}_{\gamma_2} = \left(\begin{array}{cccc}
\mathbf{1}_{2} &  &  &   \\
 & e^{4\pi\i/3} \mathbf{1}_3 &  &  \\
 &  & \mathbf{1}_{2}   &   \\
 &  &  &  e^{2\pi\i/3} \mathbf{1}_3
\end{array}\right)  \,. \label{splitbicubicWLs}
\eea
As the results on Chern-Simons invariants are usually given in terms of $SU(N)$ flat connections, it is useful to note that the above Wilson 
lines embed into an $SU(5) \subset U(5) \subset SO(10)$ subgroup of the $SO(10)$ GUT group.

Having set up the compactification, we are ready to compute the Wilson line
contribution to the superpotential.  The split-bicubic $X^{3,3}$ has both $A$-type
and $C$-type sLags.   We now turn our attention to studying these sLags in the
smooth split-bicubic quotient and computing their Chern-Simons invariants.   

\paragraph{The C-type sLags} Let us first
consider the $C$-type sLags.  The basic $C$-type sLag is obtained from the
isometric anti-holomorphic involution:
\beq
\sigma_C: \quad \zeta_j \rightarrow \bar \eta_j, \quad \eta_j \rightarrow \bar \zeta_j, \quad t_1 \rightarrow \bar t_2, \quad t_2 \rightarrow \bar t_1 \,.
\eeq
Further $C$-type sLags can be identified by considering involutions $(M,\overline{M}^{-1}) \circ \sigma_C$, and those we will consider are:
\beq
(M,\overline{M}^{-1}) \circ \sigma_C: \quad \zeta_j \rightarrow \omega^{l_j}\bar \eta_j, \quad \eta_j \rightarrow \omega^{l_j}\bar \zeta_j, \quad t_1 \rightarrow \bar t_2, \quad t_2 \rightarrow \bar t_1,
\eeq
where $l_1+l_2+l_3 = 0 \mod 3$. Together, these give three distinct $C$-type
sLags on $X^{19,19}$.

In order to understand the topology of the $C$-type sLags, it is enough to
consider the basic one.  The sLag $Q_{\sigma_C}$ can be described by the equations
\beq
0=t_1f(\zeta) - \bar t_1 \, g(\zeta)  \quad \textrm{and} \quad t_1 = \bar t_2
\eeq
in $\CP^1 \times \CP^2$.  Notice that on the sLag $t_1 = \bar t_2 \neq 0$, so
this equation reduces as a hypersurface in $\CP^2$ to:
\beq
0 = f(\zeta) -  \frac{\bar t_1}{t_1} \, g(\zeta) \,,
\eeq
which corresponds to the configuration matrix
$\left[ \CP^2 \; \vline \; 3 \right]$ describing a smooth CY 1-fold, that is,
a 2-torus.  The total sLag is then a fibration over $\RP^1$ ($t_1=\bar t_2$ in
$\CP^1$), with smooth fibers  $\T^2$.  As the monodromy of this torus bundle is
clearly trivial, the resulting 3-manifold is simply a 3-torus.  All $C$-type
sLags are diffeomorphic to the basic $C$-type sLag and hence they are also all
3-tori.

The free action of a cyclic group on a 3-torus corresponds to trivial or free
actions along each of the $S^1$ factors, so that the quotient is again a
3-torus. As explained in section \ref{CSonSeifert}, the Chern-Simons contributions from discrete Wilson lines on a 3-torus
vanish. 
 Hence the $C$-type sLags do
not contribute to the superpotential for $X^{3,3}$.

\paragraph{A-type sLags on the covering CICY}  Next we consider the $A$-type sLags, whose basic isometric anti-holomorphic
involution is:
\bea
\sigma_A: \quad \zeta_j \rightarrow \bar \zeta_j, \quad \eta_j \rightarrow \bar \eta_j, \quad t_a \rightarrow \bar t_a \,.
\eea
Further sLags can be identified from the involutions $M \circ \sigma_A$, which
we take to be:
\beq
M \circ \sigma_A:  \;\; \zeta_j \rightarrow \omega^{l_j} \bar \zeta_j, \quad \eta_j \rightarrow \omega^{m_j}\bar \eta_j, \quad t_a \rightarrow \bar t_a, \eeq
where $l_j,m_j \in \{0,1,2\}$, and $l_1+l_2+l_3 = m_1+m_2+m_3 = 0$ mod $3$.
This gives only nine $A$-type sLags in total.

The basic $A$-type sLag can be described as the complete
intersection,
\bea
0 &=& r_1f(x) - r_2g(x) = r_1 \left(x_1^3 + x_2^3 + x_3^3\right) +  r_2 x_1 x_2 x_3\nonumber \\
0 &=& r_2f(y) - r_1g(y) = r_2 \left(y_1^3 + y_2^3 + y_3^3\right) +  r_1 y_1 y_2 y_3
\eea
in $\RP^1 \times \RP^2 \times \RP^2$, with $r_a$, $x_j$ and $y_j$ being the
homogeneous coordinates on $\RP^1$, $\RP^2$ and $\RP^2$ respectively.  In
analogy with the split-bicubic itself, our real 3-manifold can then be
described as a fiber product,
\bea
&&\xymatrix{
& {Q_{\sigma_A}} \ar[dl]_\pi \ar[dr]^{\pi'} & \\
N \ar[dr]_\beta & & N' \ar[dl]^{\beta'} \\
& \RP^1 &
}
\nonumber\\
\label{pullback}
\eea
where the map $\pi$ ($\pi'$) forgets the $y_i$  ($x_i$) coordinates, and the
map $\beta$ ($\beta'$) forgets the $x_i$ ($y_i$) coordinates.

\begin{figure}[t]
\begin{center}
\includegraphics[height=15cm,angle=-90]{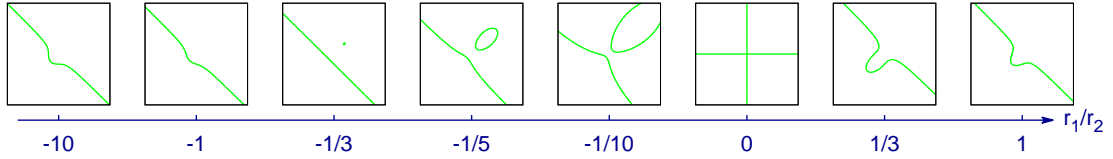}
\caption{\label{F:cubiccurves}Solutions to the cubic equation
\eqref{E:cubicplanecurve} in $\RP^2$, treating $r_1/r_2$ as a parameter.  In
the figure, we have used affine coordinates with $x_3$ scaled to unity and
plotted $x_2$ against $x_1$. The complement, $x_3 = 0$, defines an $\RP^1$
which, in the chosen affine coordinates, sits at infinity. In this way we find 
apparantly non-compact curves, but the curves that seem noncompact are
connected at infinity due to the antipodal identification on the $\RP^1$
defined by $x_3 = 0$. We see that for all $r_1/r_2\ne 0$ and $r_1/r_2\ne -1/3$
we find either a single curve which is topologically $\RP^1\cong S^1$ or a
disjoint union of two such curves. For $r_1/r_2=0$, the eq.
\eqref{E:cubicplanecurve} reduces to $x_1x_2x_3 = 0$, 
whose solution is 
three intersecting $\RP^1$'s. In this case the plot is not complete since the
entire $\RP^1$ at infinity, corresponding to $x_3=0$, is also a solution but
not shown. Finally, for $r_1/r_2=-1/3$, the solution is a disjoint union of
$\RP^1$ and a single point.}
\end{center}
\end{figure}

\begin{figure}[h]
\begin{center}
\input{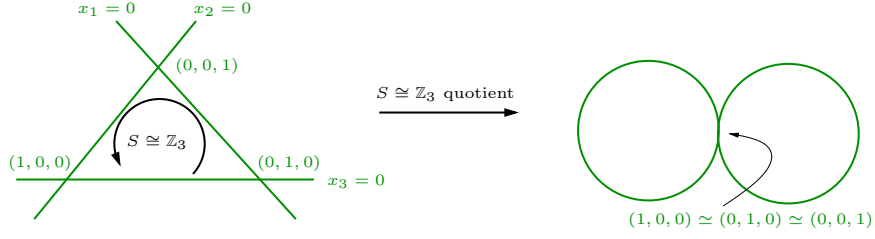}
\caption{A singular fiber in the $A$-type sLag and its quotient, solution to
the plane cubic curve  \protect\eqref{E:cubicplanecurve} at $r_1=0$. Before
modding out by $S\cong \Z_3$, it is a connected union of three $\RP^1$'s, each
two of which intersect at a point.  The Euler characteristic of this curve is
then given by $\chi(\textrm{3 intersecting } \RP^1\textrm{'s})=3\, \chi(\RP^1)
-  3\,\chi(\textrm{point}) = -3$ or $\chi(\textrm{3 intersecting }
\RP^1\textrm{'s})= b_0 - b_1 = 1 - 4 = -3$.  Modding out by the permutation
symmetry $S$, leads to a figure of eight, with Euler characteristic
$\chi(\textrm{figure of eight})=2 \chi(\RP^1) - \chi(\textrm{point}) = -1$.}
\label{F:triangle}
\end{center}
\end{figure}

In order to understand the topology of $Q_{\sigma_A}$, we start by characterizing
the topology of the 2-manifolds $N$ and $N'$, in analogy to the rational elliptic surface $dP_9$.
 $N$ is described as the
hypersurface
\beq
N=\left\{\left(r,x\right)\in \RP^1 \times \RP^2 \;\vline \;r_1 f(x) - r_2 g(x) = 0 \right\},
\eeq
and similarly for $N'$. The smooth surface $N$ can be viewed\footnote{Just as
for the complex case (see the discussion below eq. \eqref{splitbicubicmatrix}),
the manifold $N$ can also be viewed as the blowup of $\RP^2$ at three points
(where $f(x)=g(x)=0$) to $\RP^1$. This is topologically equivalent to the
connected sum of four $\RP^2$'s, i.e. a 2-sphere with four
crosscaps.  The Euler characteristic for this blowup is given by
$\chi(N)=\chi(\RP^2) - 3\,\chi(\textrm{point}) + 3\,\chi(\RP^1) = 1 -3 +
0 = -2$.} as a singular fibration over $\RP^1$ (parameterized by $r_a$) where
the fibers are given by the following cubic equation in $\RP^2$:
\beq
r_1 (x_1^3 + x_2^3 + x_3^3) + r_2 x_1 x_2 x_3 = 0~. \label{E:cubicplanecurve}
\eeq
This well-known plane cubic curve can immediately be understood with some plots, see figure \ref{F:cubiccurves}.
The generic smooth fibers are a single $\RP^1$ for $r_1/r_2 > 0$ and
$r_1/r_2 < -1/3$, or a disjoint union of two $\RP^1$'s for
$-1/3 < r_1/r_2 < 0$.\footnote{Indeed, it follows from a classic theorem due
to Harnack \cite{Harnack} that a smooth cubic in $\RP^2$ has up to two
connected components, each circles, exactly one of which must correspond to the
non-zero element of $H_1(\RP^2, \Z)\cong \Z_2$.} There are also, however, two
singular fibers:  For $r_1/r_2=-1/3$, the equation for the fiber is solved
both by the $\RP^1$ described by $x_1 = -x_2 - x_3$, and the point
$x_1= x_2=x_3$; for $r_1=0$ it gives a connected union of three $\RP^1$'s with
three singular points.
It is then straightforward to verify that the surface $N$ has Euler
characteristic (see figure \ref{F:triangle})
\[
\chi(N)=\chi(\textrm{point})\times\chi(\textrm{point}) + \chi(\textrm{point})\times\chi(\textrm{3 intersecting } \RP^1\textrm{'s}) = 1-3 = -2~,
\]
and similarly for $N'$.

\begin{figure}[h]
\begin{center}
\input{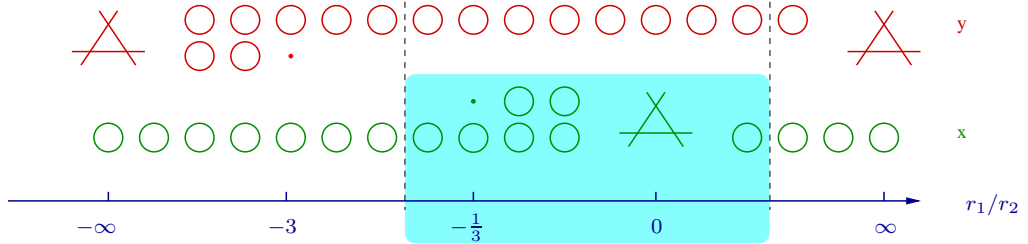}
\caption{The $A$-type sLag $Q_{\sigma_A}$ as fiber product.  The cubic curves in
the $\RP^2$ factor parameterized by $x_j$'s fibered over $\RP^1$ parameterized
by $r_a$'s give a smooth surface, $N\cong \sharp_4\RP^2$.  The same is true of
the cubic curves in $\RP^2$ parameterized by $y_j$'s fibered over $\RP^1$.
Alternatively, by cutting up the manifold into two pieces at $r_1 = \pm r_2$,
we obtain two diffeomorphic $S^1$-bundles over the bounded base $\tilde N$
indicated by the shaded area in the figure.}
\label{F:seifertfibration}.
\end{center}
\end{figure}

Building on these results, we can describe the $A$-type sLag.
First of all, we have just seen from \eqref{pullback} that it is the fiber
product $N \times_{\RP^1} N'$, i.e. a singular fibration over $\RP^1$, where the
fibers are products of two plane cubic curves described above (see eq.
\eqref{E:cubicplanecurve}).  In fact, for any ratio $r_1/r_2$ at least one of
the two plane cubic curve fibers is always a single smooth $\RP^1$ (see figure
\ref{F:seifertfibration}). By cutting up $Q_{\sigma_A}$ at two places in the
$\RP^1$ base where \emph{both} fibers are locally smooth $\RP^1$'s, say at
$r_1 = \pm r_2$, the manifold $Q_{\sigma_A}$ can be decomposed into two
diffeomorphic pieces (see figure \ref{F:seifertfibration}). We denote the piece
corresponding to $r := r_1/r_2\in [-1,1]$ by $\tilde Q_{\sigma_A}$, i.e.
\beq
\tilde Q_{\sigma_A} = \left\{( r,x,y)\in [-1,1]\times \RP^2\times \RP^2\,\vline\,  rf(x) - g(x) = 0 = f(y) - rg(y)\right\}.
\eeq
Since the fibers above $r=\pm 1$ are 2-tori, the above cutting operation is an
example of a torus decomposition, which we discussed in section
\ref{CSonSeifert}.
The map $\tilde\pi:\tilde Q_{\sigma_A}\to \tilde N$, where
$\tilde \pi(r,x,y) = (r,x)$ and
\beq
\tilde N = \left\{(r,x)\in [-1,1]\times \RP^2\,\vline\, rf(x) - g(x) = 0\right\},
\eeq
defines an $S^1$-bundle over $\tilde N$ since $\tilde \pi$ projects out
\emph{smooth} $S^1$ fibers (see figure \ref{F:seifertfibration}),
\beq
\tilde \pi^{-1}(r,x) = \left\{y\in \RP^2\,\vline\, f(y) - rg(y) = 0\right\}\cong \RP^1\cong S^1.
\eeq
This is a trivial Seifert fibration (i.e. $S^1$-bundle over a smooth surface,
$\tilde N$), where the base $\tilde N$ has two circular boundaries.

\paragraph{A-type sLags on the quotient}  
Up to now, we have identified the $A$-type sLags in the simply connected
split-bicubic, $X^{19,19}$, together with their topological structure.
Next we have to understand how the sLags are modified when we mod out
$X^{19,19}$ by the discrete symmetry $\Gamma=S\times R$ to obtain $X^{3,3}$. The only
$A$-type sLag on $X^{3,3}$ that can inherit a Wilson line is the basic one,
which may only inherit a Wilson line associated with $S$. In the covering
space $X^{19,19}$, the permutation group does not act on the base $\RP^1$ of the
sLag $Q_{\sigma_A}$. Therefore, the quotient sLag
$Q_{\sigma_A}/S\cong Q_{\sigma_A}/\Z_3$ can still be described as a fibration over
$\RP^1$ with the fibers being a product of two plane cubic curves
\eqref{E:cubicplanecurve} subject to identifications. Let us consider the
action of $S$ on these plane cubic curves. We first note that $S$ is a
symmetry of the defining polynomial \eqref{E:cubicplanecurve} so that for a
fixed $r=r_1/r_2$ each
plane cubic curve is mapped to itself by $S$. Moreover the only fixed point of
$S$ in the ambient $\RP^2$ is $x_1=x_2=x_3$. We now examine how the permutation
group $S$ acts on the four topologically different
types of plane cubic curve (see figure \ref{F:cubiccurves}).  
Referring to \protect\eqref{E:cubicplanecurve}:
\begin{itemize}
\item For $r= -1/3$, the plane cubic curve is topologically a disjoint union
of a circle and the point $x_1=x_2=x_3$. The permutation $S$ acts freely on the
circle component which thus stays topologically a circle
after modding out by $S$ and the point component is a fixed point.
\item For $r \notin [-1/3,0]$, the plane cubic curve is topologically a single
circle which is mapped freely to itself by $S$. Again, the quotient curve
remains a circle.
\item For $r = 0$, the plane cubic curve consists of three intersecting
circles as depicted in figure \ref{F:triangle}. Each circle is given by the
vanishing of one of the coordinates, and hence the permutation
action maps the circles onto one another. Moreover on each circle there are
two distinguished points that map into each other, namely the intersection
points of that circle with the other two. The quotient
topology is then easily verified to be the so-called figure of eight.
\item For $r \in (-1/3,0)$, the plane cubic curve consists of two disjoint
circles. The permutation group $S$ acts freely within each circle component.
This can be seen as follows, one of the two circles
has all $x_j$ with the same sign (the smaller circle in the corresponding diagrams of
figure \ref{F:cubiccurves}) while the $x_j$ in the other circle do not have the same sign.
\end{itemize}

\begin{figure}[h]
\begin{center}
\input{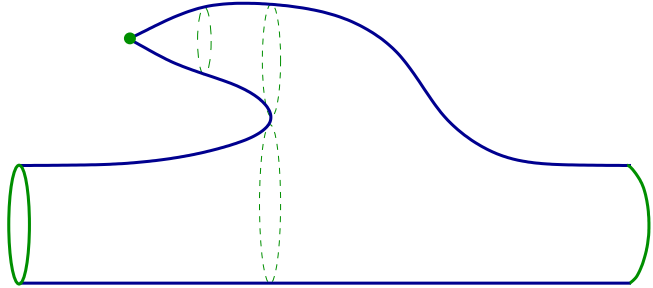}
\caption{The base $\tilde N/\Z_3$ of the quotient sLag $Q_{\sigma_A}/\Z_3$ after
torus decomposition.}
\label{F:moddedoutbase}
\end{center}
\end{figure}

\begin{figure}[h!]
\centering
\input{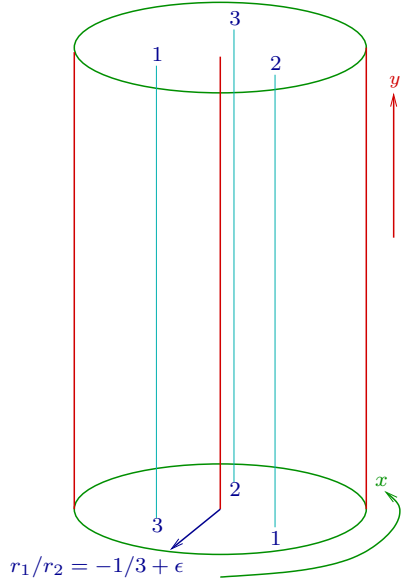}
\caption{The exceptional fiber in the Seifert fibration of the quotient
$\tilde Q_{\sigma_A}/\Z_3$.  The exceptional fiber lies above the fixed point of
the $\Z_3$ action in the orbit surface $\tilde B$. The figure shows the 
structure close to an exceptional fiber as follows.  We consider a disk
neighbourhood of the orbifold point $(\tilde r,\tilde x_1/\tilde x_3,\tilde
x_2/\tilde x_3) = (-1/3,1,1)$ in the base surface. The disk forms the base of
a fibered solid torus, which is the product
$D^2_{\tilde r,\tilde x} \times I_{\tilde y}$ with the ends of the interval
$I_{\tilde y}$ identified after twisting by an angle of $2\pi/3$.  The center of
the disc $\left\{0\right\}$ lifts to the core circle of the solid torus, and
points in $D^2_{r,x} - \left\{0\right\}$ lift to fibers that wrap 3 times around
the core in the longitudinal direction and 1 times in the meridianal
direction. An example of a fiber is shown in blue, the three line segments are
joint together as indicated when the endcaps of the
cylinder are glued together.  Thus the data describing the exceptional fiber
is $(p, q)=(3,1)$ or $(\alpha, \beta)=(3,1)$ .}
\label{F:singseifert}
\end{figure}

As $S$ acts trivially on the base $\RP^1$ parameterized by $r_a$, we can now
perform essentially the same torus decomposition as for the unquotiented sLag,
namely cut $Q_{\sigma_A}/\Z_3$ along toroidal boundaries located at
$r_1 = \pm r_2$. Each of the two resulting components is now diffeomorphic to
$\tilde Q_{\sigma_A}/\Z_3$. Before we mod out by $S$, $\tilde Q_{\sigma_A}$ is a
$S^1$-bundle
over the smooth
base $\tilde N$. The permutation group $S\cong \Z_3$ acts freely within each
$S^1$-fiber so that the quotient $\tilde Q_{\sigma_A}/\Z_3$ is also an
$S^1$-bundle, but over the base manifold $\tilde N/\Z_3$. As explained above,
$\tilde N$ has precisely one fixed point located at
$(r,x_1,x_2,x_3) = (-1/3,1,1,1)$. Increasing $r$ from $r=-1/3$ to
$r=-1/3+\epsilon$, the isolated fixed point grows into a circle (see
figure \ref{F:seifertfibration}) so
that the coordinates $r$ and $x$ locally parameterize a disk neighbourhood of
the fixed point. The permutation group $S\cong \Z_3$ acts on this disk
neighbourhood by rotating the disk about the fixed point
in its center. It is therefore clear that $\tilde N/\Z_3$ has an orbifold
singularity of order three at the center of the disk whereas everywhere else
the quotient $\tilde N/\Z_3$ is smooth. Thus the space $\tilde Q_{\sigma_A}/\Z_3$
is now a non-trivial Seifert fibration with one exceptional fiber, see figure
\ref{F:singseifert}. The manifold has Seifert invariant (c.f. \eqref{Sfinv}):
\beq
\tilde Q_{\sigma_A}/\Z_3=(O,o,0;0,(3,1))\,,
\eeq
where we have used that the underlying topology of the orbit surface is a
cylinder (see figure \ref{F:moddedoutbase}) and 
recalled that the section obstruction
$b$ is trivial on manifolds with boundary. 

\paragraph{Wilson lines on the A-type sLags and their Chern-Simons invariants}  Given the Seifert invariant, one can immediately write down a
presentation of the fundamental group (c.f. \eqref{Sfpi1}):
\beq \label{pi1QA}
\pi_1(\tilde Q_{\sigma_A}/\Z_3) = \langle h,c_0,c_1,d_1,d_2 \; \vline \;  h\text{ is central},\, c_0 = c_1^3h = c_0c_1d_1d_2 = 1 \rangle\,.
\eeq
This fundamental group is infinite and non-Abelian.


This particular fundamental group together with the appropriate gluing
condition to compose $Q_{\sigma_A}/\Z_3= \tilde{Q}_{\sigma_A}/\Z_3 \cup \tilde{Q}_{\sigma_A}/\Z_3$, does not allow one to define a $\Z_3$ Wilson line consistently on
the entire sLag $Q_{\sigma_A}/\Z_3$. This is explained in appendix \ref{A:CSSB}.
We can therefore conclude that the corresponding Chern-Simons invariant vanishes
\beq
CS(A, Q_{\sigma_A}/\Z_3) = 0\,,
\eeq

\paragraph{A basis for the third homology group and the flux superpotential}
Finally, we should check whether or not we span the basis for the third homology group, as required to obtain all the Wilson line contributions 
to the Chern-Simons flux superpotential.  This is described in more detail in the appendix \ref{S:splitbicubic}.
The rank of the $A$- and $C$-type intersection matrix can be computed to be zero for the smooth split-bicubic\footnote{Note that this does not 
imply that all the $A$-type and $C$-type sLags are homologically equivalent, but only that the number of linearly independent homology elements 
covered by the cycles is \emph{at least} zero.}.  However, the singular split-bicubic, with complex structure parameters $a=b=c=0$ has additional 
$A$-type and $C$-type sLags, due to its larger set of isometric anti-holomorphic involutions.  
Starting from this singular limit -- and deferring certain subtleties in the intersection theory in that limit -- we can obtain a 
set of deformed sLags, which complete a basis for the third homology group of the smooth quotient split-bicubic.  We have to consider the Wilson 
lines and Chern-Simons invariants for these deformed sLags which complete the basis.  Whether or not Wilson lines wrap the cycles can be inferred 
from the singular limit, where it is clear from section \ref{S:WLonsLags} that Wilson lines can project non-trivially on the basic $A$-type sLag 
and $C$-type sLags.  All the $C$-type sLags in the singular limit of the split-bicubic are smooth, and they are
topologically 3-tori. Hence, like the basic $C$-type sLag, their Chern-Simons
invariants are zero.  Recalling that the basic $A$-type sLag also has a
vanishing Chern-Simons invariant, we therefore conclude that all the
Chern-Simons invariants vanish and we can write down the full Wilson line
contribution to the Chern-Simons flux superpotential,
\beq
W_\text{CS}= 0\,.
\eeq 
In contrast to the quintic, one therefore cannot introduce fractional terms in the
flux superpotential coming from the visible or hidden sector Wilson lines.  On one hand the consistency of the leading order 10D supersymmetric CY compactification is clear, and on the other hand Chern-Simons fluxes from Wilson lines cannot help with moduli stabilization.

\section{Conclusions}
\label{S:conclusions}

Discrete Wilson lines are a key ingredient in heterotic Standard Model constructions based on Calabi-Yau compactifications.\footnote{See \cite{Blumenhagen:2006ux} for some Standard Model like constructions without Wilson lines on simply-connected Calabi-Yaus.} 

They are introduced to break grand unified gauge groups down to the standard model whilst maintaining supersymmetry and the control that this provides. 
However, they can sometimes induce a non-trivial fractional $H$-flux via their Chern-Simons contributions, which may affect the internal 
self-consistency of the assumed string background and could lead to possibly unintended phenomenological consequences such as high-scale 
supersymmetry breaking. Since, for a given Wilson line, the presence or absence of fractional $H$-flux is not a choice, it is important to 
develop methods for its computation.

We analysed this problem for complete intersection Calabi-Yau manifolds that admit freely acting symmetry groups of discrete rotations, 
$R$, and cyclic permutations, $S$. We used  the well understood special Lagrangian submanifolds based on isometric anti-holomorphic 
involutions as explicit representatives for the 3-cycles of the third homology group. If they span a basis for the third homology group, the 
full background superpotenial from Chern-Simons flux can be expressed in terms of Chern-Simons invariants on these submanifolds. The special 
Lagrangian submanifolds come in two types, the $A$-type associated with complex conjugation of the coordinates in the ambient projective spaces, 
and the $C$-type associated with complex conjugation and exchange of coordinates between any two of the ambient projective spaces of equal dimension. 
In a systematic analysis we determined which sLags could potentially inherit non-trivial Wilson lines from the Calabi-Yau space. This first step is 
model independent.

The actual value of the Chern-Simons invariant depends both on the topology of the submanifold and the choice of Wilson line, but it is 
computable on a model-by-model basis. As an illustration we carried out this computation for two explicit complete intersection Calabi-Yaus, 
namely for the quintic and the split-bicubic. The 3-dimensional spaces we encountered in these models are Seifert fibered 3-manifolds or composition thereof.  
For Wilson lines in such spaces we can compute the Chern-Simons invariants by applying results from the mathematics literature.

For the quintic modded out by $\Z_5 \times \Z_5$, we were able to obtain an expression for the full superpotential induced by Wilson lines.   
The result depends on whether we choose to embed the Wilson line in the $R$ or $S$ factor of the Calabi-Yau fundamental group.  Notice that
the low energy particle spectrum and couplings are independent of this choice. 
Choosing an $R$-type Wilson line, all Chern-Simons invariants and the superpotential are vanishing in this model.  In this way, we can ensure a 
consistent leading order supersymmetric Calabi-Yau 10D compactification.  Choosing an $S$-type Wilson line, by contrast, there is a non-vanishing 
Chern-Simons invariant and superpotential, which might be used for moduli stabilization, but may also introduces subtleties regarding the self-consistency 
of the string background.

We then progressed to the potentially realistic three generation quotient split-bicubic with two discrete Wilson lines.  The special Lagrangian 
submanifolds we found for the smooth quotient split-bicubic do not generate the full third homology group, but by starting from a more symmetric singular limit, we potentially
identified deformed sLags that do span a basis.
Contrary to the quintic case we found that the Wilson lines do not generate any $H$-flux and therefore do not contribute to the 
flux superpotential. This is completely independent of the choice of Wilson lines and is due solely to the topological properties of 
the three dimensional submanifolds in the split-bicubic. This is a very interesting result, since it  supports the self-consistency of 
the models constructed on the split-bicubic, but it also 
means that moduli stabilization must be achieved by some mechanism different to the one proposed in \cite{Gukov:2003cy}, 
see e.g. \cite{Anderson:2011cza, Anderson:2011ty, Cicoli:2013rwa, Anguelova:2010ed, Anguelova:2010qd}. 

Our work leaves several important open questions.  The consistency of incorporating Chern-Simons flux into supersymmetric Calabi-Yau 
compactifications with gaugino condensation has not yet been established.  
In any case, ultimately, it would be necessary to compute the Chern-Simons flux (and its superpotential) from Wilson lines in 
any explicit Calabi-Yau compactification.  Our procedure should be applicable to a wide range of models, but there are also some model dependent steps. 
It would be invaluable to develop methods to implement these within computerized scans like \cite{Anderson:2013xka}. 
Finally, it would be important to check for global worldsheet anomalies due to Wilson lines in explicit models.  

\section*{Acknowledgments}
We are grateful to Lara Anderson, Gilberto Bini, Volker Braun, Athanasios Chatzistavrakidis, James Gray, Alexander Haupt, 
Andrea Jaramillo Puentes, Malek Joumaah, Diego Matessi, Nicola Pagani, Eran Palti, Frithjof Schulze, Eric Sharpe, Inken Siemon and 
Davide Veniani for helpful conversations. 
FA would like to thank Virginia Tech for hospitality during the final stages of this work.    
FFG would like to thank the SCGP for hospitality and financial support during
the 2014 Simons Summer Workshop.  This work was in part supported by the German Research Foundation (DFG) within the Cluster 
of Excellence QUEST and the RTG 1463 "Analysis, Geometry and String Theory".

\appendix
\section{The intersection matrix for sLags}\label{S:appendix}
In this appendix we give details on how the intersection matrices of sLags are calculated. A more detailed discussion is presented in \cite{Brunner:1999jqpa,Palti:2009bt,Denef:2001ix}.
\subsection{The quintic}
For the quintic we use the simplest polynomial \eqref{fermatquintic}
\beq\label{A:fermatquintic}
\sum_{i=0}^5 z_i^5 = 0,
\eeq
where $z_i\in\CP^4$. From the definitions of involutions presented in section \ref{involutions} we notice that the only possible involutions we can consider are of $A$-type. We will limit ourselves to $A$-type sLags defined as the simultaneous solutions of \eqref{A:fermatquintic} and
\[
z_i = \omega^{l_i}\bar{z}_i,
\]
where $\omega = \e^{2\pi \i/5}$ and $l_i\in\Z^5$. The topology of the sLags is well known to be $\RP^3$. The intersection number of two sLags is given by the Euler number of the intersection subspace \cite{Brunner:1999jqpa,Palti:2009bt}. For instance in the quintic, the subspace is given by the solution to
\[
z_i = \omega^{l_i}\bar{z}_i,\qquad z_i = \omega^{k_i}\bar{z}_i,
\]
together with the quintic equation, \eqref{A:fermatquintic}. The dimension of the intersection is
\[
3 - n\,,
\]
where $n$ is the number of $l_i\neq k_i$. For example if $l_i=k_i$ for all $i$, the intersection is simply the sLag itself which is three dimensional. If $k_5 = 1$ and all other $k_i$'s and $l_i$'s are zero then $z_1$ simultaneously has to satisfy $z_5 = \bar{z}_5$ and $z_5 = \omega \bar{z}_5$ which implies that $z_5=0$. We therefore lose one degree of freedom and the intersection is a surface, as is consistent with $n=1$. The Euler number of the surface is $1$, because surface intersections of a pair of manifolds, each diffeomorphic to $\RP^3$, is topologically a $\RP^2$. This can also be noted from the fact that the intersection is a single solution of a real equation in $\RP^3$. The intersection number in this case is $-1$, where the sign is due to an orientation between the sLags. The orientation can be calculated from
\[
\sgn \prod_i \sin \f{2\pi (l_i - k_i)}{5},
\]
where only non-trivial terms are included in the product \cite{Brunner:1999jqpa,Palti:2009bt,Denef:2001ix}. In summary if $n$ is odd then the intersection number is equal to $\pm 1$, where the sign is determined by the orientation. If $n$ is even, then either the intersection is the sLag itself or a curve, topologically a circle. In both cases the intersection number vanishes.

It is convenient to introduce the notation
\[
\langle k_1k_2k_3k_4k_5 | l_1l_2l_3l_4l_5 \rangle,
\]
to denote the intersection matrix. From the above example we see that
\[
\langle 00001|00000 \rangle = -1.
\]
The orientation formula, together with the fact that intersection numbers with $n$ even vanish, ensures that the intersection matrix is anti-symmetric.

The sLags defined by the rotation angles $l_i$ are not all independent. By employing the scaling symmetry $z_i \mapsto \e^{\pi i \lambda/5} z_i$ we effectively transform the $l_i$'s by the formula $l_i \mapsto l_i + \lambda$ for $\lambda \in \Z_5$. We have only used the scaling symmetry to make this transformation and so the two sLags have to be the same. We therefore define an equivalence class
\[
[l_i] \equiv \{ l_i \sim l_i + \lambda,\ \forall \lambda \in \Z_5\}.
\]
We calculate the intersection number of two equivalence classes simply by summing the intersection numbers of all elements in the classes
\[
\langle [k_i] | [l_i] \rangle \equiv \sum_{k_i\in[k_i],l_i\in[l_i]} \langle k_1k_2k_3k_4k_5  | l_1l_2l_3l_4l_5  \rangle.
\]
This does not give the actual numerical value for the intersection number, but the whole intersection matrix is scaled by a common factor which of course does not affect its rank.
We also want to compute the intersection matrix of a CICY which is modded out by a discrete group. This modding out is taken care of in the same way as for the scaling symmetries. The equivalence classes of sLags are enlarged by the discrete symmetry. For example in the quintic we mod out by $\Z_5$ generated by the cyclic permutation $z_i\mapsto z_{i+1}$ which translates to a permutation of the $l_i$'s, $p:l_i \mapsto l_{i-1}$. We then define a new equivalence class
\[
[l_i]_{\Z_5} \equiv \{ l_i \sim l_i + \lambda,\ l_i \sim p^\kappa(l)_i=l_{i-\kappa}, \ \forall \lambda,\kappa \in \Z_5\},
\]
and again the intersection number of equivalence classes is defined by the sum
\[
\langle [k_i]_{\Z_5} | [l_i]_{\Z_5} \rangle \equiv \sum_{k_i\in[k_i]_{\Z_5},l_i\in[l_i]_{\Z_5}} \langle k_1k_2k_3k_4k_5  | l_1l_2l_3l_4l_5  \rangle.
\]

Using this procedure we find that the rank of the intersection matrix precisely matches the dimension of the third homology group of the quintic and the modded out quintic.

\subsection{The split-bicubic}\label{S:splitbicubic}
For the split-bicubic a similar procedure to that used for the quintic holds.   We identify sLags using isometric antiholomorphic involutions of the CICY.  Then, using the description of these sLags as complete intersections, we can easily compute their intersection loci, the corresponding Euler characteristics and hence the intersection numbers.  Taking care of the orientations and the scaling symmetry as done for the quintic, we can then compute the rank of the intersection matrix.  We will, however, encounter one additional complication, which is that we must pass through a singular limit of the split-bicubic in order to find sufficient 3-cycles to span a basis of the third homology group.

Ensuring first a choice of complex structure parameters that give a smooth CY ($a=b=0, \;c\neq 0$), we take:
\bea\label{A:splitbicubic}
P_1(t,\zeta) = t_1(\zeta_1^3+\zeta_2^3+\zeta_3^3) + c \, t_2\zeta_1\zeta_2\zeta_3\,,\nonumber\\
P_2(t,\eta) = t_2(\eta_1^3+\eta_2^3+\eta_3^3) + c\, t_1\eta_1\eta_2\eta_3\,.
\eea
As discussed in the main text, this smooth split bicubic has 9 $A$-type sLags and 3 $C$-type sLags, described respectively by $(k_1,k_2,k_3)$ with $k_1+k_2+k_3=0\mod 3$ and $(k_1,k_2,k_3,l_1,l_2,l_3)$ with $k_1 + k_2 + k_3 = l_1 + l_2 + l_3 \mod 3 = 0\mod 3$, where we have taken $c=\epsilon$ real.  Notice that, as we will discuss further below, more sLags could be obtained by taking the singular CY with $a=b=c=0$, indeed it is then easy to identify 81 $A$-type sLags and 9 $C$-type sLag. Also, different sets of  9 $A$-type and 3 $C$-type sLags can be obtained by choosing different smooth choices for $c$, $c = \epsilon \omega^n$ with $\omega=e^{2\pi\i/3}$ and $n=0,1,2$.  These are labelled by $(k_1,k_2,k_3)$ with $k_1+k_2+k_3= 2n \mod 3$ and $(k_1,k_2,k_3,l_1,l_2,l_3)$ with $k_1 + k_2 + k_3 = l_1 + l_2 + l_3 \mod 3 = 2n \mod 3$.  The equations describing these sLags as complete intersections in $\RP^1\times \RP^2 \times \RP^2$ are identical for all $A$-type sLags and all $C$-type sLags.

In \ref{intersnum} we present the intersection numbers for all $A$- and $C$-type sLags in the unmodded smooth split-bicubic, given by the Euler characteristic of the intersection loci.
\begin{table}
\centering
\begin{tabular}{l|rrr}
Intersection & A $\cdot$ A & A $\cdot$ C & C $\cdot$ C\\
\hline
point & 1&1&0\\
curve & 0&0&0\\
surface&-2&0&0
\end{tabular}
\caption{\label{intersnum}The intersection numbers for intersections of $A$- and $C$-type sLags in the split bicubic, given by the Euler characteristic of the intersection loci.}
\end{table}
The only non-trivial entry in table \ref{intersnum} is the surface intersection of two $A$-type sLags, so let us explain how this can be obtained. An $A$-type sLag is given by the solution of
\[
\zeta_i = \omega^{l_i}\bar{\zeta}_i,\qquad \eta_i = \omega^{k_i}\bar{\eta}_i,\qquad t_i = \bar{t}_i,
\]
together with the defining polynomials \eqref{A:splitbicubic}.
For two such sLags, a simultaneous solution is a surface when only one of the angles $k_i$ and $l_i$ are different. Let us then consider the basic $A$-type sLag with $k_i=l_i=0$ intersecting with the sLag defined by $k_1=1$ and other $k$'s and $l$'s vanishing. We find that the intersection locus is defined by $\zeta_1 = 0$ and $\zeta_2,\zeta_3,\eta_i$ and $t_i$ real. We can denote $\zeta_i = x_i$, $\eta_j=y_j$ and $t_i=r_i$ to distinguish from the complex coordinates on the ambient space. The intersection surface satisfies the equations
\[
0 = r_1(x_2^3 + x_3^3) = r_2(y_1^3 + y_2^3 + y_3^3) + r_1y_1y_2y_3,
\]
where $r, (x_2,x_3)\in \RP^1$ and $y\in\RP^2$. As indicated in the table \ref{intersnum}, this surface has Euler characteristic $-2$. We  
can  see this by the fact that for $r_1\ne 0$ the first equation simply has a point solution $x_2=-x_3$, the second equation, has a solution 
space which is topologically a $\RP^1\cong S^1$ except for $r_2 =0$ and $r_1=-3r_2$. For $r_2=0$ the solution space is three intersecting 
$\RP^1$'s and for $r_1=-3r_2$ the solution space is a point and a $\RP^1$. The total Euler characteristic of the surface is determined only 
by these contributions, i.e. $\chi = -3 +1 = -2$ where $-3$ is the Euler characteristic of the three intersecting $\RP^1$'s.

Computing finally the intersection matrix, it turns out to be the zero matrix.  A similar computation can be carried out for the modded out 
split bicubic but of course the rank of the intersection matrix in all cases turns out to vanish.  
Note that this does not imply that all the $A$-type and $C$-type sLags are homologically equivalent, but only that the number of linearly 
independent homology elements covered by the cycles is \emph{at least} zero.

We can, deferring certain subtleties to be stated below, identify a set of deformed sLags which do span a basis for the third homology group of the smooth split-bicubic.  We do so by considering first the singular split-bicubic, taking $a=b=c=0$:
\bea\label{A:singsplitbicubic}
P_1(t,\zeta) = t_1(\zeta_1^3+\zeta_2^3+\zeta_3^3)\,,\nonumber\\
P_2(t,\eta) = t_2(\eta_1^3+\eta_2^3+\eta_3^3)\,.
\eea
We can fill out an intersection matrix for this CICY as 
follows\footnote{An essential notion in intersection theory is to be able to move cycles using equivalence relations (in our case homological equivalence) 
to ensure that they are in a generic position, whereby the intersection product of two subvarieties consists of their set-theoretic intersection. 
For singular spaces, this may not always be possible.  We will proceed by assuming that the cycles considered are in sufficiently generic positions.  
Proving that this is so, however, is a difficult mathematical question beyond the scope of this work.}.
First note that it is easy to write 
down equations describing all 81 $A$-type sLags and 9 $C$-type sLags, as well as identify point, curve and surface intersections as described 
above.  Note that the sLags and the surface intersections are singular, but also that each intersection of a given dimension is described by the same equation.  
Next, observe that 9 out of the 81 $A$-type sLags and 3 out of the 9 $C$-type sLags persist as sLags when we deform to a smooth CICY, taking $c$ from 0 to $\epsilon$.  
In going to this smooth limit, we can use the result that the intersection number for smooth sLags is given by the Euler number of the 
intersection\footnote{This result follows from the isomorphism between the tangent bundle and normal bundle for Lagrangian manifolds. 
As the self-intersection number of a manifold $X$ is $X.X=e(NX)[X]$, we then have $X.X=e(NX[X])=e(TX[X])=\chi[X]$.}. 
Assuming that the
intersection numbers do not change in going back to the singular limit -- which may not be justified -- they are given by table \ref{intersnum}.  Moreover, these are the 
intersection numbers for all point, curve and surface intersections, given that they are described by the same equations.  Having filled out 
the intersection matrix, we can compute its rank,
finding 16 and 8, respectively, for $X^{7,7}=X^{19,19}/S$ and $X^{3,3}=X^{19,19}/S\times R$.  That is, the $A$-type and $C$-type sLags span the 
basis for the third homology group of the singular CICY.  Finally, we know that all these sLags survive as 3-cycles when we deform to a smooth CICY\footnote{Indeed, for many kinds of singularities, the map between third homology groups $H_3(X_{smth}) \rightarrow H_3(X_{sing})$ is 
surjective, so that cycles can disappear when going to the singular limit, but no new cycles can appear.  One way to see this in our case is 
to notice that we can define the holomorphic 3-form and the periods in the singular limit, and deform them away from the singular limit.  
Therefore, the cycles also exist in the smooth limit.  We thank Volker Braun for explaining this to us.}, even though they are not all fixed 
point sets of any isometric antiholomorphic involution (and thus likely not all sLags).  In this way, we obtain a set of deformed sLags that 
generate the full third homology group of
the smooth (quotient) split bicubic. The topology of 27 out of the 81 deformed $A$-type sLags and all deformed $C$-type sLags are the same as 
that of the basic $A$-type and $C$-type sLags, as can be seen by considering the different smooth limits, 
$c= \epsilon, \epsilon \omega, \epsilon \omega^2$ which are diffeomorphic to each other.

Whilst a mathematically rigorous computation of the intersection matrix for the singular CICY is beyond the scope of this paper, 
the final matrix ranks obtained might be considered compelling indicators that the subtleties mentioned can be overcome.

\section{Chern-Simons invariant on the basic A-type sLag of the split-bicubic} \label{A:CSSB}
In this appendix we compute the Chern-Simons invariant of the sLag $Q_{\sigma_A}/S \cong Q_{\sigma_A}/\Z_3  $ in the quotient 
split-bicubic, $X^{3,3}$. To do so, we first have to understand how the Wilson line associated with the symmetry group $S\cong \Z_3$, 
which is a homomorphism $\rho:\; \pi_1(X^{3,3}) \rightarrow SO(10)$, is compatible with the fundamental group $\pi_1(Q_{\sigma_A}/\Z_3)$ 
of the sLag. In fact, we will show that the Wilson line associated with $S$ on $X^{3,3}$ cannot project to a Wilson line on the sLag 
$Q_{\sigma_A}/\Z_3$.

The strategy is to check whether the fundamental group of the manifold $Q_{\sigma_A}/\Z_3$ admits a homomorphism 
$\rho:\; \pi_1(Q_{\sigma_A}/\Z_3) \rightarrow SO(10)$ whose image can be written as \eqref{splitbicubicWLs}. We start by recalling that 
the sLag has been cut into two pieces, $\tilde{Q}^{(I)}_{\sigma_A}/\Z_3$ with $I=1,2$, as in figure \ref{F:seifertfibration}.   Each piece 
is a Seifert fibered manifold with boundary and their fundamental group is given by \eqref{pi1QA}.  In order to understand the generators 
of the fundamental group, we look at the fibration structure of the manifold described in section \ref{SS:splitbicubic}, and list the 
non-contractible loops present:
\begin{itemize}
\item $h^{(I)}$ is associated with the $S^1$ fiber;
\item $c_0^{(I)}$ is associated with an eventual twisting of the base $\tilde{N}^I$;
\item $c_1^{(I)}$ corresponds to the non-contractible loop around the orbifold point in $\tilde{N}^I$;
\item $d_1^{(I)}, d_2^{(I)},$ are the two boundaries of the cylinder $\tilde{N}^I$, see figure \ref{F:moddedoutbase}.
\end{itemize}
The next step is to glue the two manifolds $\tilde{Q}^{(1)}_{\sigma_A}/\Z_3$ and $\tilde{Q}^{(2)}_{\sigma_A}/\Z_3$ along the two boundaries 
given by the plane cubic curves at the points $r=r_1/r_2=\pm1 $.  As we have already seen, the boundaries are 2-tori, and the gluing condition 
is an automorphism of the torus, namely an $SL(2,\Z)$ transformation, that maps the two circular boundaries of $\tilde{Q}^{(1)}_{\sigma_A}/\Z_3$ 
to the ones of $\tilde{Q}^{(2)}_{\sigma_A}/\Z_3$ (and the reverse for the other boundary). Note that the symmetry group $S\cong \Z_3$ acts such 
that there is no twisting of the two fibers in the neighbourhood of $r=r_1/r_2=\pm1 $ on the original uncut manifold, where we recall that the 
fibers are given by the two plane cubic curves (see figures \ref{F:seifertfibration} and \ref{F:cubiccurves}). Therefore, we can write the gluing 
conditions as follows.  Along the boundary $r=1$ we have
\begin{eqnarray}
&&h^{(1)}=d^{(2)}_1, \label{br1.1}\\
&&d^{(1)}_2=h^{(2)},  \label{br1.2}
\end{eqnarray}
and along the boundary $r=-1$ we have
\begin{eqnarray}
&&h^{(2)}=d^{(1)}_1, \label{br2.1}\\
&&d^{(2)}_2=h^{(1)}. \label{br2.2}
\end{eqnarray}
So far, together with the relations in \eqref{pi1QA}, we have listed all the topological ingredients of our sLag $Q_{\sigma_A}/\Z_3$. Wilson 
lines on the sLag would correspond to the homomorphism $\pi_1(Q_{\sigma_A}/\Z_3) \rightarrow SO(10)$, given by:
\beq \label{expmap}
\rho: h^{(I)} \mapsto e^{2\pi i Y^{(I)}}; \qquad \rho: c^{(I)}_k \mapsto e^{2\pi i X^{(I)}_k}, \;\; k=0,1;\qquad  \rho: d^{(I)}_l \mapsto e^{2\pi i D^{(I)}_l}, \;\; l=1,2;
\eeq
where at least one of the generators of the fundamental group should generate a $\Z_3$ subgroup, in order to be mapped to the 
matrices in \eqref{splitbicubicWLs}. To check if this is possible we start from the relations (in \eqref{pi1QA}) given by
\begin{eqnarray}
\left(c_1^{(I)}\right)^3 h &=& 1 \;\Rightarrow\; (3 X^{(I)}_1 + Y^{I} )\in {\rm diag}(\Z), \label{pi1rel1}\\
c_0^{(I)}\left(h^{(I)}\right)^b &=& 1 \; \Rightarrow\; ( X^{(I)}_0 + b Y^{I} )\in {\rm diag}(\Z), \label{pi1rel2}\\
 c_0^{(I)} c_1^{(I)} d_1^{(I)} d_2^{(I)} &=& 1 \;\Rightarrow\; (X^{(I)}_0+ X^{(I)}_1 +  D^{(I)}_1 +  X^{(I)}_2) \in {\rm diag}(\Z), \label{pi1rel3}
\end{eqnarray}
where we have used \eqref{expmap} and ${\rm diag}(\Z)$ is the set of integer valued 10$\times$10 diagonal matrices. Again using the map  
$\rho$ in \eqref{expmap}, the boundary gluing conditions (\ref{br1.1}-\ref{br1.2}) become
\begin{eqnarray}
&&Y^{(1)}=D^{(2)}_1\, \mod\, {\rm diag}(\Z),\label{BR1.1}\\
&&D^{(1)}_2=Y^{(2)} \, \mod\, {\rm diag}(\Z). \label{BR1.2}
\end{eqnarray}
and (\ref{br2.1}--\ref{br2.2}) become
\begin{eqnarray}
&&Y^{(2)}=D^{(1)}_1 \, \mod\, {\rm diag}(\Z), \label{BR2.1}\\
&&D^{(2)}_2=Y^{(1)} \, \mod\, {\rm diag}(\Z). \label{BR2.2}
\end{eqnarray}
Since we want a Wilson line that is a homomorphism $\rho$ of $\Z_3$ into $SO(10)$, suppose that every generator $g$ fulfils the following relation
\beq
g^3=1.
\eeq
This implies that $3X^{(I)}_1 \in {\rm diag}(\Z)$, which together with \eqref{pi1rel1} gives also $Y^{(I)}\in {\rm diag}(\Z)$. Plugging these 
results into \eqref{pi1rel2}, we find that also $X^{(I)}_0\in {\rm diag}(\Z)$. Using now the boundary gluing conditions (\ref{BR1.1}-\ref{BR2.2}) 
and, plugging them into \eqref{pi1rel3}, we obtain that also $X^{(I)}_1 \in  {\rm diag}(\Z)$. To sum up, we have obtained a completely trivial 
representation, and therefore $\Z_3$ Wilson lines do not project onto the sLag $Q_{\sigma_A}/\Z_3$.

\bibliographystyle{utphys}
\bibliography{refs}
\end{document}